\documentclass[reqno]{amsart}
\usepackage{tikz-cd}
\usetikzlibrary{graphs,quotes}
\usepackage{amssymb}
\usepackage{amsthm}
\usepackage{pgfplots}
\usepackage[a4paper, total={6in, 9in}]{geometry}
\usepackage{mathtools}
\usepackage{ stmaryrd }
\usepackage{cite} 
\usepackage{mathabx}
\usepackage{amssymb}
\usepackage{cite}
\usepackage{mathabx} % Adds a lot of math symbols and modifies the common ones
\usepackage[colorlinks]{hyperref}
\usepackage{hyperref} % Turns \ref and \cite command into hyperlinks
\usepackage[utf8]{inputenc}
\usepackage{enumerate}
\usepackage{amsmath}
\usetikzlibrary{math}
\usepackage{amssymb}
\usepackage[T1]{fontenc} % Allows diacritics to be copied from the pdf properly
\usepackage[utf8]{inputenc} % Allows diacritics to be included in the source

\usepackage{color,soul}
\definecolor{ItalianApricot}{rgb}{1,0.7,0.5}
\sethlcolor{ItalianApricot}

\theoremstyle{plain}
\newtheorem{thm}{Theorem}[section]

\newtheorem{lem}[thm]{Lemma}

\theoremstyle{definition}
\newtheorem{defn}[thm]{Definition}

\theoremstyle{remark}
\newtheorem{remark}[thm]{Remark}
\newtheorem{oq}[thm]{Open question}

\numberwithin{equation}{section}

\renewcommand{\phi}{\varphi}
\newcommand{\f}{\mathcal{F}}

\renewcommand{\epsilon}{\zeta}
\renewcommand{\phi}{\varphi}

\newcommand{\tr}{\text{Tr}}
\newcommand{\MLR}{\text{MLR}}
\newcommand{\oln}{\text{ln}}
\newcommand{\C}{\mathbb{C}}

\begin{document}
 
\title[]{von Neumann Entropy and Quantum Algorithmic Randomness}
\author[]{Tejas Bhojraj}

\date{\today}% It is always \today, today,
 
\address[Tejas Bhojraj]{School of Computer Science, University of Auckland, NZ}
\email{tejasbhojrajpnp@gmail.com}

% \subjclass[2010]{Primary ***; Secondary ***, ***}
%\input{title.tex}

\maketitle

\begin{abstract}
A state $\rho=(\rho_n)_{n=1}^{\infty}$ is a sequence such that $\rho_n$ is a density matrix on $n$ qubits. It formalizes the notion of an infinite sequence of qubits. The von Neumann entropy $H(d)$ of a density matrix $d$ is the Shannon entropy of its eigenvalue  distribution. We show: (1) If $\rho$ is a computable quantum Schnorr random state then $\lim_n [H(\rho_n )/n] = 1$. (2) We define quantum s-tests for $s\in [0,1]$, show that $\liminf_n [H(\rho_n)/n]\geq \{ s: \rho$ is covered by a quantum s-test $\}$ for computable $\rho$ and construct states where this inequality is an equality. (3) If $\exists c \exists^\infty n H(\rho_n)> n-c$ then $\rho$ is strong quantum random. Strong quantum randomness is a randomness notion which implies quantum Schnorr randomness relativized to any oracle.
(4) A computable state $(\rho_n)_{n=1}^{\infty}$ is quantum Schnorr random iff the family of distributions of the $\rho_n$'s is uniformly integrable. We show that the implications in (1) and (3) are strict. 
\end{abstract}

% \subjclass[2010]{Primary ***; Secondary ***, ***} 
\tableofcontents
\section{Introduction}

We study the initial segment von Neumann entropy of states (infinite sequences of qubits). Quantum Schnorr randomness for computable states is shown to strictly imply an entropy rate of 1. We define quantum s-tests and show that if $s$ is greater than the entropy rate of a state then the state is covered by a quantum s-test. Having infinitely many initial segments with maximum entropy (upto a constant) is shown to strictly imply strong quantum randomness. Quantum Schnorr randomness of computable states is shown to be equivalent to the uniform integrability of a family of functions given by the eigenvalues of the initial segments. We use general information theoretic arguments which may be of independent interest.
\subsection{Classical and quantum algorithmic randomness}
The theories of computation and of information have been generalized to the quantum setting  \cite{bernstvaz, Berthiaume:2001:QKC:2942985.2943376, nielsen2010quantum, Mller2007QuantumKC, PeterGacs_2001}. Algorithmic randomness, an area using notions from computation and information, has recently been extended to the quantum realm\cite{unpublished, bhojraj2020quantum, qpl, Bhojraj:2021vko,  Bhojraj2021PrefixfreeQK}.

Algorithmic randomness studies the complexity of elements of Cantor space: the space of bitstrings equipped with the uniform measure and a compact topology. Consider the bitstring $1010\cdots$ which has an easily describable `pattern' to it. By contrast, the bitstring obtained by tossing a fair coin repeatedly is almost always `patternless'. A central theme in algorithmic randomness is to quantify the sense in which the later bitstring is more `random' or `irregular' than the former (See the textbooks \cite{misc, misc1, DBLP:series/txtcs/Calude02, BookE} for more detail). Martin-L{\"o}f and Schnorr randomness are important notions of algorithmic randomness.

Quantum algorithmic randomness, an extension of algorithmic randomness to the quantum setting, studies the randomness of \emph{qubitstrings} (infinite sequences of qubits) and was introduced by Nies and Scholz \cite{unpublished}. They formalized a notion of an infinite sequence of qubits, referred to it as a state and introduced quantum Martin-L{\"o}f randomness and quantum Solovay randomness for states. Quantum Schnorr randomness and weak quantum Solovay randomness were defined in \cite{bhojraj2020quantum}.

Quantum algorithmic randomness has nice parallels with classical algorithmic randomness. Martin-L{\"o}f randomness for elements of Cantor space is equivalent to quantum  Martin-L{\"o}f randomness of the states they induce. Thus, quantum  Martin-L{\"o}f randomness generalizes Martin-L{\"o}f randomness. Similarly to the classical setting, quantum Martin-L{\"o}f randomness is equivalent to quantum Solovay randomness and is strictly stronger than quantum Schnorr randomness\cite{bhojraj2020quantum}. A quantum version of the law of large numbers is satisfied by quantum Schnorr random states\cite{bhojraj2020quantum}. A quantum analogue (QK) of the prefix-free Kolmogorov complexity (K) has been developed \cite{Bhojraj2021PrefixfreeQK}.

Quantum Schnorr randomness has an initial-segment complexity based characterization which mirrors the initial-segment complexity characterization of Schnorr randomness using computable measure machines\cite{Bhojraj2021PrefixfreeQK}. All notions of initial segment complexity for states developed so far use Turing machines to compress the initial segments: \cite{unpublished} uses unitary machines, a type of quantum Turing machines while \cite{Bhojraj2021PrefixfreeQK} uses classical prefix-free Turing machines.

\subsection{The von Neumann entropy}
Take a discrete probability distribution $P=(p_1,\cdots, p_d)$. Its Shannon entropy ($H(P)$) defined as
$H(P):=-\sum_{i=1}^d p_i \text{log}_2 (p_i)$ is a foundational notion in information theory.

A finite qubitstring is mathematically described by a density matrix (a Hermitian, positive semidefinite matrix with trace one). A density matrix's eigenvalues form a distribution on the set of its eigenvectors: the probability assigned to an eigenvector is its eigenvalue. 
A density matrix's von Neumann entropy is the Shannon entropy of the distribution given by its eigenvalues.\footnote{ The von Neumann entropy plays a pivotal role in quantum coding theory, quantum entanglement, quantum Kolmogorov complexity, quantum statistical mechanics and in quantum information generally \cite{smb, Epstein2021OnTA, epstein:hal-04072076 ,Berthiaume:2001:QKC:2942985.2943376, Mller2007QuantumKC, wilde_2013, nielsen2010quantum}.}  We denote the von Neumann entropy of a density matrix $d$ by $H(d)$. As the Shannon entropy is higher for more `uniform' distributions, $H(d)$ is high when $d$'s eigenvalue distribution is uniformly spread out over its eigenvectors and is small if the distribution peaks at a `few' eigenvectors.

\subsection{Motivation}  A state $\rho=(\rho_n)_{n=1}^{\infty}$ is a coherent sequence such that $\rho_n$ is a density matrix on $\C^{2^n}$ for all $n$.\footnote{The precise meaning of `coherent' will be defined later.} The length $n$ initial segment of $\rho$ is a length $n$ qubitstring described by $\rho_n$. We informally sketch the intuition which led us to explore the connection between the initial segment von Neumann entropy of a state and its randomness.

Let $d$ be a density matrix and $S$ be a subspace which is the range of  a Hermitian projection, $G$. We informally refer to the quantity $\tr(dG)$ as the `projection of $d$ onto $S$'.

Roughly, a quantum Schnorr test is a computable sequence of subspaces of small dimension. A state `fails' the test if its initial segments have large projections onto infinitely many subspaces of the test. A state is not quantum Schnorr random if it fails some test.

A small $H(\rho_n)$ implies that $\rho_n$'s distribution is far from uniform and concentrates at a few eigenvectors. These eigenvectors span a small dimensional subspace onto which $\rho_n$ has a large projection. If all $\rho_n$'s have small $H(\rho_n)$ then such small dimensional subspaces exist for each $\rho_n$. Further, if $\rho$ is computable then these subspaces can be computed and form a test which $\rho$ fails. This intuition underlies Section \ref{liminf} where we show that for computable $\rho$, $\liminf_n H(\rho_n)/n < 1$ implies that $\rho$ is not quantum Schnorr random.

As the computability of $\rho$ is essential to computing the small dimensional subspaces of the test, the above heuristic fails to apply to non-computable states. In fact, there are non-computable, quantum Schnorr random states with zero initial segment entropy (See Section \ref{liminf}).

It is also the case that `high' initial segment entropy implies randomness. An initial segment with high entropy has a nearly uniform distribution. I.e., its eigenvalues are spread out evenly over its eigenvectors. The singular value decomposition then implies that there is no small dimensional subspace onto which the initial segment has a large projection.\footnote{
The singular value decomposition implies that if the rank of $G$ is $k$ and $\tr(dG)>\delta$ then the sum of the first $k$ many largest eigenvalues of $d$ exceeds $\delta$ (See Theorem \ref{thm:10}).} This implies that $\rho$ does not fail a test if the $\rho_n$'s have high entropy. This underlies the ideas in Section \ref{ls} where we show that if there is a $c$ such that the entropy of $\rho_n$ exceeds $n-c$ infinitely often, then $\rho$ is strongly quantum random.

The above ideas are borne out by examples: The state $(2^{-n}I_n)_n$ (where $I_n$ is the $2^n$ dimensional identity matrix) is quantum Schnorr random by definition and has maximum initial segment entropy. In Lemma 3.5 in \cite{qpl}, we construct a computable state $\rho=(\rho_n)_n$  which is not quantum Martin-L{\"o}f random and which has low initial segment entropy in that $n-H(\rho_n)$ tends to infinity. 

Sections \ref{liminf} and \ref{ls} indicate that the initial segment entropy is a measure of the complexity of states. The initial segment Kolmogorov complexity ($C$ and $K$) is a measure of complexity of infinite bitstrings \cite{misc, misc1}. It thus seems that Kolmogorov complexity and entropy play similar roles in quantum and classical algorithmic randomness respectively. Section \ref{rqr} is motivated by this similarity. Solovay s-tests are related to the initial segment Kolmogorov complexity of bitstrings: $\liminf_n K(X\upharpoonright n)/n = \liminf_n C(X\upharpoonright n)/n  = \inf \{s: X$ is covered by a Solovay s-test $\}$. 
We define quantum s-tests and show that: $\liminf_n H(\rho_n)/n \geq \inf \{s: \rho $ is covered by a quantum s-test $\}$ for a computable $\rho$.

The results of Sections \ref{liminf},\ref{rqr} and \ref{ls} suggest that the randomness of a computable state $\rho=(\rho_n)_n$ is related to the limiting behavior of the  distributions of the $\rho_n$'s as $n$ grows. Instead, one may ask if there is any \emph{global} property of the entire \emph{family} of these distribution which correlates with randomness. Section \ref{ui} shows that uniform integrability (See page 133 in \cite{rudin}), is one such. 

For a state $\rho=(\rho_n)_n$ we define a family $\f^\rho = (f^\rho_n)_n$ of functions on $[0,1)$ such that $f^\rho_n$ encodes the distribution of $\rho_n$. The family is constructed such that for any $r$ there is an interval $J\subseteq [0,1)$ such that $\tr(\rho_n G) \leq \int_J f^\rho_n $ for any rank $r$ projection $G$. This uses Theorem \ref{thm:10}, a corollary of the singular value decomposition. Since the quantum Schnorr randomness of $\rho$ depends on the behavior of $\tr(\rho_n G)$ for various computable projections $G$ of small rank, this result can be used to show that quantum Schnorr randomness is equivalent to the uniform integrability of $\f^\rho$ for computable $\rho$.

\subsection{Overview and results}
We state the relevant preliminaries in Section \ref{sec:4}. We give an overview of the main results contained in the Sections.

\begin{itemize}
    \item Section \ref{liminf} shows that if $\rho=(\rho_n)_n$ is a computable quantum Schnorr random then $H(\rho):=\liminf_n H(\rho_n )/n = 1$. This implication is strict.
    \item Section \ref{general} proves a technical information theoretic result used in Sections \ref{liminf} and \ref{rqr}.
    \item Section \ref{rqr} defines quantum s-tests and shows that $H(\rho) \geq \inf \{ s: \rho $ is covered by a quantum s-test $\}$ for computable $\rho$. The quantum typical subspaces theorem is used to construct states $\rho$ such that $H(\rho) = \inf \{ s: \rho $ is covered by a quantum s-test $\}$.
    \item Section \ref{svd} proves a consequence of the singular value decomposition used in the next two sections.
    \item Section \ref{ls} defines strong quantum randomness, a randomness notion implying quantum Schnorr randomness relativized to any oracle. We show that $\rho$ is strong quantum random if there exists a $c>0$ such that for infinitely many $n$, $H(\rho_n)> n-c$. Also, this implication is strict.
    \item Section \ref{ui} shows that for a computable $\rho$, the family of functions $\f^\rho$ is uniformly integrable if and only if $\rho$ is quantum Schnorr random.
\end{itemize}

\section{Preliminaries}
\label{sec:4}
The set of non-negative integers is denoted by $\omega$. Cantor space (See Section 1.8 \cite{misc}) will be denoted by $2^\omega$. The set of finite bitstrings is denoted by $2^{<\omega}$. The set of bitstrings of length $n$ is denoted by $2^n$. For any $\sigma \in 2^{\omega}$, we use $\llbracket \sigma \rrbracket$ to denote $\{X \in 2^\omega: X \succ \sigma \}$. For a $A \subseteq 2^{\omega}$, $\llbracket A \rrbracket := \bigcup_{\sigma \in A} \llbracket \sigma \rrbracket$. Lebesgue measure will be denoted by $\mu$. `log' will denote the logarithm to the base 2. Let $I_k$ denote the identity matrix on $\mathbb{C}^{2^k}$. $I_1$ will be denoted by $I$.
\subsection{Quantum information}
As per the bra-ket notation, a vector $v \in \mathbb{C}^m$ is denoted as $|v\rangle$ and its dual is denoted by $\langle v|$. In a fixed basis of $\mathbb{C}^m$, $|v\rangle$ is a column vector while $\langle v|$ is a row vector which is equal to $((|v\rangle)^*)^T $, the complex conjugate transpose of $|v\rangle$ (See 2.1.1 and 2.1.4 in \cite{nielsen2010quantum}). A Hermitian projection, $H$ is a self-adjoint projection. Recall that $\tr(H)$ equals the rank of $H$ and the dimension of the range of $H$. By `projection', we will always mean a Hermitian projection.

\begin{defn}
\label{dens}
A density matrix on $\C^{2^n}$ is a Hermitian, positive semidefinite matrix with trace 1. It  describes a $n$-qubit quantum system and has an orthonormal eigenbasis with its eigenvalues lying in $[0,1]$ and summing to one (See Section 2.4.1 in \cite{nielsen2010quantum}).
Let $d$ be a density matrix on $\C^{2^n}$ and let it have an orthonormal eigenbasis $(|\psi_{i}\big> )_{i\leq 2^n}$ and corresponding eigenvalues $(\alpha_{i})_{i\leq 2^n}$.  So, \[ d = \sum_{i\leq 2^n} \alpha_{i}|\psi_{i}\big> \big< \psi_{i}|. \] 

The eigenvalues $(\alpha_{i})_{i\leq 2^n}$ give a probability distribution on the set $(|\psi_{i}\big> )_{i\leq 2^n}$, where $|\psi_{i}\big>$ is assigned probability $\alpha_i$.
The von-Neumann entropy of $d$ (See Section 11.3 in \cite{nielsen2010quantum}), denoted $H(d)$, is defined to be the Shannon entropy of the discrete probability distribution $(\alpha_i)_{i\leq 2^n}$: \[H(d) = -\sum_{i\leq 2^n}\alpha_{i}\text{log}_{2}(\alpha_{i}).\]

We will refer to  probability distribution $(\alpha_i)_{i\leq 2^n}$ as the `distribution of $d$'. Recall that $H(d)\leq n$ and $H(d)=n$ when $d=2^{-n}I_n$. As the Shannon entropy is larger for more `uniform' distributions, $H(d)$ is high when $d$'s distribution is evenly spread out over its eigenvectors and is small if the distribution peaks at a `few' eigenvectors.
\end{defn}
\subsection{Classical and Quantum Algorithmic randomness}
\label{qardefn}

Many authors have come up with the notion of an infinite sequence of qubits (a state) \cite{unpublished,smb,brudno}. We use the one given by Nies and Scholz \cite{unpublished}. 
\begin{defn}
\label{state}
A state, $\rho=(\rho_n)_{n=1}^\infty$ is an 
infinite sequence of density matrices such that $\rho_{n} \in \mathbb{C}^{2^{n} \times 2^{n}}$ and $\forall n>1$,  $PT_{\mathbb{C}^{2}}(\rho_n)=\rho_{n-1}$. Here, $PT_{\mathbb{C}^{2}}$ denotes the partial trace which `traces out' the last qubit from $\mathbb{C}^{2^n}$ (See Section 2.4.3 in \cite{nielsen2010quantum}).
\end{defn}
The state $\rho$ represents an infinite sequence of qubits whose first $n$ qubits are described by $\rho_n$. The definition requires $\rho$ to be coherent in the sense that for all $n$, $\rho_n$, when `restricted' via the partial trace to its first $n-1$ qubits, has the same measurement statistics as the state on $n-1$ qubits given by $\rho_{n-1}$. The following state will be the quantum analogue of Lebesgue measure.
\begin{defn}
\label{def:tr}
Let $\tau=(\tau_n)_{n=1}^\infty$ be the state given by setting $\tau_n = \otimes_{i=1}^n I$. This is called the tracial state. This state has maximal initial segment von Neumann entropy as $H(\tau_n)=n$ .
\end{defn}

\begin{defn}
\label{defn:sigclass}

If $\rho$ is a state and $G$ is a projection on $\mathbb{C}^{2^n}$ then $\rho(G):= \tr (\rho_{n}G)$. Note that $\tau(G)=2^{-n}(\tr(G))= 2^{-n}$(rank($G$)).
\end{defn}

\begin{defn}
A special projection is a Hermitian projection matrix with complex algebraic entries. 
\end{defn}

We now consider computable sequences of special projections. 
A sequence $(a_n)_{n=1}^\infty$ of natural numbers is said to be computable if there is a computable function $\phi$ such that $\phi(n)=a_n$.

A complex algebraic  number is the root of a polynomial with rational coefficients. By a result of Rabin, there is a  1-1 function from the complex algebraic numbers to $\omega$ such that the field operations in the image are computable. This function may be used to identify a special projection with a natural number (See pg.8 in \cite{unpublished}).  We define a sequence of special projections to be computable when the corresponding sequence of natural numbers is computable. 

\begin{defn} 
\label{compstate}
A state  $\rho = (\rho_n)_{n=1}^\infty$ is computable if each $\rho_n$ is a special projection and the sequence $(\rho_n)_n$ is computable. 
\end{defn}
A clopen subset $C \subseteq 2^\omega$ is a set such that there is a finite set, $F\subset 2^{<\omega}$ such that $C= \llbracket F \rrbracket$. The quantum analogue of a clopen subset of $2^\omega$ is a special projection (See Defn.3.2 in \cite{unpublished} and the discussion following it). A finite total Solovay test is a sequence $(\llbracket F_n \rrbracket)_n$ of clopen sets sets such that $(F_n)_n$ is a computable sequence of finite subsets of $2^{<\omega}$ and such that $\sum_n \mu(C_n)$ is a computable real number (See Defn.7.2.21 in \cite{misc1}). An element of $2^\omega$ is Schnorr random iff it passes all finite total Solovay tests (See 7.2.22 in \cite{misc1}) where $X$ is said to pass $(C_n)_n$ if $X\in C_n$ for at most finitely many $n$. This motivated our definition of quantum Schnorr randomness  (Defn.2.6 in \cite{bhojraj2020quantum}):
\begin{defn}

A quantum Schnorr test (q-S test) is a computable sequence of special projections, $(S^{m})_{m=1}^\infty$ such that $\sum_{m}\tau(S^{m}) $ is a computable real number\cite{bhojraj2020quantum}. A state $\rho$ fails the q-S test $(S^{m})_{m}$ at order $\delta>0$ if $\rho(S^{m})>\delta$ for infinitely many $m$. A state $\rho$ passes the q-S test $(S^{m})_{m}$ at order $\delta>0$ if it does not fail it at $\delta$. A state is quantum Schnorr random if it passes all q-S tests at all $\delta >0$.
\end{defn}

\begin{remark}
\label{schnorr}

Recall that a real number $r\in [0,1]$ is said to be computable if there is a computable function, $\phi$ from the natural numbers to the rationals such that $|r-\phi(n)|\leq 2^{-n}$ for all $n$. 

Let $(S^{m})_{m=1}^\infty$ be a computable sequence of special projections, such that $\tau(S^{m})\leq 2^{-m}$. Then the function $\phi$ where $\phi(n) := \sum_{m=1}^n \tau(S^m) $ is a computable function from naturals to rationals. Further, \[\bigg|\sum_{m=1}^\infty  \tau(S^m) - \phi(n)\bigg|=\bigg|\sum_{m=1}^\infty  \tau(S^m) - \sum_{m=1}^n \tau(S^m)\bigg| =\sum_{m=n+1}^\infty \tau(S^m) \leq \sum_{m=n+1}^\infty 2^{-m}\leq 2^{-n}. \]
This shows that a computable sequence of special projections $(S^{m})_{m=1}^\infty$ such that $\tau(S^{m})\leq 2^{-m}$ is a q-S test.

\end{remark}

\section{Quantum Schnorr randomness and entropy rate.}
\label{liminf}

If $\rho=(\rho_n)_n$ is a state then $ {H(\rho_{n})}/{n}$ can be interpreted as the entropy `per qubit' of the first $n$ qubits of $\rho$. The quantity $\lim_{n}(H(\rho_{n})/n)$, when the limit exists, is the asymptotic per-qubit entropy of $\rho$ and is similar to the entropy rate of a stochastic process (See Section 4.2 in \cite{cover2012elements}.)
Theorem \ref{thm:99} shows that computable quantum Schnorr random states have maximum entropy rate.

Recall from \ref{dens} that the eigenvalues of a density matrix define a distribution on its eigenvectors where the probability assigned to an eigenvector is its eigenvalue.
Let $d$ be a computable density matrix with small $H(d)$. Then $d$'s distribution is non-uniform and peaks at a few eigenvectors. Let $G$ be the projection onto this small dimensional subspace spanned by these few eigenvectors. Then $G$ has small rank and $\tr(Gd)$, the total probability assigned by $d$'s distribution to these few eigenvectors, is large. Further, $G$ is computable as $d$ is.

The main theme of this section is to extend this idea to computable states. We show that if $\rho=(\rho_n)_n$  is computable and $\liminf_n H(\rho_n)/n < 1$ then there is a computable sequence of small rank projections $(G_n)_n$ and $\delta>0$ such that $\exists^\infty \tr(\rho_n G_n)>\delta$. For all $n$, $G_n$ is the small rank projection obtained by letting $d$ be $\rho_n$ in the idea above.

The sequence $(G_n)_n$ is computable as the eigendecomposition of $\rho_n$ is uniformly computability in $n$. Since the $G_n$'s have small rank, $(G_n)_n$ gives a q-S test which $\rho$ fails. We thus conclude that if a computable state $\rho=(\rho_n)_n$ is quantum Schnorr random then  $\liminf_n H(\rho_n)/n = 1$.\footnote{Another perspective on the above heuristic is: If $\rho $ is a computable quantum Schnorr random, the quantum Schnorr randomness of $\rho$ cannot stem from the eigenvectors of the $\rho_n$'s as they are uniformly computable in $n$ and hence non-random. Rather, the randomness of $\rho$ should be a result of the distributions of the $\rho_n$'s being highly uniform over its eigenvectors and hence impossible to capture using the small rank projections in a q-S test. The distributions of the $\rho_n$'s being highly uniform implies that $\liminf_n H(\rho_n)/n$ is large.}

The computability of $\rho$ is crucial in the above idea. Although a non-computable $\rho$ may have $\liminf_n {H(\rho_n)}/{n} < 1$, the few eigenvectors at which the distribution is concentrated, need not be computable. This precludes the construction of a q-S test which $\rho$ fails. E.g., let $Z\in 2^\omega$ be Schnorr random (and thus uncomputable). The state $\rho = (\rho_n)_n$ where $\rho_n = |Z \upharpoonright n \rangle \langle Z \upharpoonright n|$ is quantum Schnorr random (See Lemma 3.9 in \cite{bhojraj2020quantum}). However, $H(\rho_n)/n =0$ for all $n$. We begin with an important definition:
\begin{defn}
    \label{hrate}
    Let $\rho=(\rho_{n})_{n}$ be a state. Then $H(\rho) := \liminf_{n} H(\rho_{n})/n$. 
\end{defn}

\begin{thm}
\label{thm:99} 
Let $\rho=(\rho_{n})_{n}$ be a computable, quantum Schnorr random state. Then \[ \lim_{n}\dfrac{H(\rho_{n})}{n}  = 1.\] 
\end{thm}

\begin{proof}
Note that $\limsup_{n} (H(\rho_{n})/n)\leq 1$. 
It thus suffices to show that $ H(\rho)  = 1$. 

\emph{Proof sketch:} Suppose towards a contradiction that $H(\rho)<1$. Fix a $\theta$ such that $H(\rho) < \theta<1$. This implies by Lemma \ref{lemma} that there is a $\delta >0$ such that for infinitely many $n$, the sum of the $2^{n \theta }$ many largest eigenvalues of $\rho_n$ exceeds $\delta$.

Since $\rho$ is computable, these eigenvalues and their corresponding eigenvectors can be computed uniformly in $n$. This allows us to build a q-S test which $\rho$ fails at $\delta$.

\emph{Proof details:} 
Towards a contradiction, let $\rho=(\rho_{n})_{n}$ be a computable quantum Schnorr random state with $H(\rho)<1$. Let $\theta$ be such that $H(\rho)<\theta <1$. Each $\rho_n$ can be expressed as (See Definition \ref{dens}), \[\rho_{n} = \sum_{i\leq 2^n} \alpha^{n}_{i}|\psi^{n}_{i}\big> \big< \psi^{n}_{i}|.
\]
We assume that the eigenvalues are labelled such that $\alpha^{n}_{1} \geq \alpha^{n}_{2}\dots \geq\alpha_{2^{n}}^{n}$.

We use Lemma \ref{lemmagene} which is restated here:

\begin{lem}
\label{lemma}
Let $\rho=(\rho_{n})_{n}$ be a state with $H(\rho)<1$. For each $n$ let $\rho_{n} = \sum_{i\leq 2^n} \alpha^{n}_{i}|\psi^{n}_{i}\big> \big< \psi^{n}_{i}|$ where the eigenvalues are labelled such that $\alpha^{n}_{1} \geq \alpha^{n}_{2}\dots \geq\alpha_{2^{n}}^{n}$. Let $\epsilon$ be arbitrary such that $H(\rho)<\epsilon<1$. Then
there is a $\delta>0$ such that $\sum_{i \leq \lceil2^{n\epsilon}\rceil}\alpha^{n}_{i} > \delta$  for infinitely many $n$. 
\end{lem}

Fix the $\delta>0$ given by Lemma \ref{lemmagene} applied to $\epsilon := \theta$ and to $\rho$. The $\delta$ may be assumed to be rational. The lemma gives that a constant ($\delta$) amount of distribution of $\rho_n $ concentrates at the first $\lceil 2^{n \theta} \rceil$ many largest eigenvalues, infinitely often. 
We use this to define a q-S test which $\rho$ fails at order $\delta$. 
We describe the construction of a special projection $G^{m}$ uniformly in $m$. $(G^m)_m$ will be the desired q-S test.

\emph{Construction of $G^m$:}
Find a $n$ such that both of the following hold:
\begin{enumerate}

    \item $\sum_{i \leq \lceil2^{n\theta}\rceil}\alpha^{n}_{i} > \delta$.
    
    \item $\dfrac{2^{n\theta}+1}{2^n}<2^{-m}$. 
\end{enumerate}    
Find $  \sum_{i\leq 2^{n}} \alpha^n_i |\psi^{n}_{i}\big> \big< \psi^{n}_{i}|,$ the eigendecomposition of $\rho_n$
and set
\[ G^{m} := \sum_{i\leq \lceil 2^{n\theta}\rceil} |\psi^{n}_{i}\big> \big< \psi^{n}_{i}|. \] 

Let $n_m$ denote the $n$ found above.

\emph{Verification:}

By Lemma \ref{lemma}, there are infinitely many $n$ such that condition (1) holds. Further, as $\rho$ is computable, the satisfaction of condition (1) for a given $n$ can be computably checked. For any $m$, condition (2) holds for almost every $n$ since $\theta<1$. This shows that $n_m$ can be computed uniformly in $m$. The computability of $\rho$ implies that $G^{m}$ can be computed uniformly in $m$ and hence that $(G^m)_m$ is a computable sequence of special projections. Note that $\tr(G^m) \leq 2^{n\theta}+1$. Since $n_m$ satisfies condition (2) we have that $\tau(G^m)= 2^{-n_m}\tr(G^{m})<2^{-m}$. This shows that  $(G^{m})_m$ is a q-S test by Remark \ref{schnorr}. Since $n_m$ satisfies condition (1) we have that $\tr(\rho_{n_m}G^{m}) = \sum_{i \leq \lceil2^{n_m \theta}\rceil}\alpha^{n_m}_{i}>\delta$. So $\rho(G^m)>\delta$ (See Definition \ref{defn:sigclass}). 

As this holds for all $m$, $\rho$ fails the q-S test $(G^m)_m$ at order $\delta$. Thus $\rho$ is not quantum Schnorr random giving the desired contradiction.
\end{proof}

\begin{remark}
\label{mlrtoo}
    
We remark that since quantum Martin-L{\"o}f randomness implies quantum Schnorr randomness, Theorem \ref{thm:99} also shows that quantum Martin-L{\"o}f random states have maximum entropy rate.
\end{remark}

We now construct a computable state $\rho=(\rho_n)_n$ which is not quantum Schnorr random and such that $\lim_n H(\rho_n)/n = 1$. This shows that the implication in Theorem \ref{thm:99} is strict.
\begin{lem}
\label{ex}
    There is a computable state, $\rho$ such that $\lim_n H(\rho_n)/n =1$ and $\rho$ is not quantum Schnorr random.
\end{lem}
\begin{proof}
Let $d_0 := |0\rangle\langle 0|$ and for $i>0$, let
$d_i := |0\rangle\langle 0| \otimes 2^{-i}I_i$.
So $d_i$ is a density matrix on $i+1$ qubits.
The informal idea is to let $\rho$ be `$ \bigotimes_{i=1}^\infty d_i$'. The formal definition is:  Let $\xi(m):= m+\sum_{i=1}^m i $. Define $\rho=(\rho_n)_n$ to be the state such that 
\[\rho_{\xi(m)} = \bigotimes_{i=1}^m d_i,\]
for all $m$. If $n$ is not of the form $\xi(k)$ for some $k$ then let $m$ be the greatest number such that $\xi(m)<n$ and let

\[\rho_n = \rho_{\xi(m)}\otimes d_{n-\xi(m)-1}.\]

We first show that:

\emph{The state $\rho$ is not quantum Schnorr random:}
For all $m$, let $G^m$ be the following special projection, \[G^m := \bigotimes_{i=1}^m |0\rangle\langle 0| \otimes I_i = \bigotimes_{i=1}^m 2^i d_i.\]

The sequence $(G^m)_m$ is uniformly computable in $m$.
Note that $G^m$ is on $\mathbb{C}^{2^{\xi(m)}}$ and that \[\tr(G^m)= \prod_{i=1}^m 2^i = 2^{\sum_{i=1}^m i} = 2^{\xi(m) - m}.\]  So $\tau(G^m)=2^{-\xi(m)}\tr(G^m)=2^{-m}$, showing that $(G^m)_m$ is a q-S test by Remark \ref{schnorr}.

Further, $\rho$ fails this test at order 1 since for any $m$, we have that $ \rho(G^m)= \tr(\rho_{\xi(m)} G^m)=1$.
We now to show that: \emph{The state $\rho$ has maximum entropy per qubit.}
Note that
\[H(\rho_{\xi(m)})=\sum_{i=1}^m H(d_i)=\sum_{i=1}^m i = \xi(m)-m.\]
So, 
\begin{align}
\label{xim}
    \lim_m \dfrac{H(\rho_{\xi(m)})}{\xi(m)}=\lim_m \dfrac{\xi(m)-m}{\xi(m)}= 1 - \lim_m \dfrac{m}{\xi(m)}= 1.
\end{align}

Suppose now that $n$ is not of the form $\xi(k)$ for any $k$. Let $m$ be the greatest $k$ such that $\xi(k)<n$. Then,
\[H(\rho_n)=H(\rho_{\xi(m)})+H(d_{n-\xi(m)-1})=\xi(m)-m + n - \xi(m)-1 = n-m-1 > \xi(m) - (m+1).\]
Using this and that $n < \xi(m+1)$, 
\[\dfrac{H(\rho_{n})}{n}> \dfrac{\xi(m)-(m+1)}{\xi(m+1)}.\]
As the $m$ in the above depends on $n$, we denote it $m_n$ instead of $m$. Then, $m_n$ tends to infinity as $n$ tends to infinity. So, taking the limit as $n$ ranges over numbers not of the form $\xi(k)$ for some $k$,
\begin{align}
\label{normal}
    \lim_n \dfrac{H(\rho_{n})}{n} \geq  \lim_n\dfrac{\xi(m_n)-(m_n +1)}{\xi(m_n +1)} = \lim_n\dfrac{\xi(m_n)}{\xi(m_n +1)}= \lim_n\dfrac{\xi(m_n + 1)- (m_n +2)}{\xi(m_n +1)}=1.
\end{align}

By \ref{xim} and \ref{normal}, when we take the limit over all $n$, we get that,
\[\lim_n \dfrac{H(\rho_{n})}{n}=1.\]

\end{proof}

\section{An information theoretic lemma}
\label{general}
We prove a general   information theoretic lemma about the entropy rate of arbitrary, not necessarily computable, states. This is applied to computable states in Sections \ref{liminf} and \ref{rqr}.
\begin{lem}
\label{lemmagene}
Let $\rho=(\rho_{n})_{n}$ be a state with $H(\rho)<1$. For each $n$ let $\rho_{n} = \sum_{i\leq 2^n} \alpha^{n}_{i}|\psi^{n}_{i}\big> \big< \psi^{n}_{i}|$ where the eigenvalues are labelled such that $\alpha^{n}_{1} \geq \alpha^{n}_{2}\dots \geq\alpha_{2^{n}}^{n}$.

Let $\epsilon$ be arbitrary such that $H(\rho)<\epsilon<1$. Then
there is a $\delta>0$ such that for infinitely many $n$, \[\sum_{i \leq \lceil2^{n\epsilon}\rceil}\alpha^{n}_{i} > \delta. \]
I.e., 
$\lim_n \sum_{i \leq \lceil2^{n\epsilon}\rceil}\alpha^{n}_{i}$,
if it exists, is not equal to zero. 
\end{lem}

The lemma says that a constant ($\delta$) amount of distribution of $\rho_n $ concentrates at the first $\lceil 2^{n \epsilon} \rceil$ many largest eigenvalues, infinitely often. 
The proof is by contraposition: Assuming that no such $\delta$ with the properties above exists allows us to bound the $H(\rho_n)$'s from below and show that $H(\rho) \geq \epsilon$.

\begin{proof}

Suppose towards a contradiction that for all $ \delta$ there exists $ N_{\delta}$ such that,  \[ n>N_{\delta} \Rightarrow \sum_{i \leq \lceil2^{n\epsilon}\rceil}\alpha^{n}_{i} \leq \delta.\]
Fix a $\delta< 0.5$ and a $n>N_{\delta}$. For this $n$, define the sequence, $(r^{n}_{i})_{i\leq 2^{n}}$ as follows. The sequence will be defined piecewise in three pieces
\begin{enumerate}
    \item For $i\leq \lceil2^{n\epsilon}\rceil$, let $r^{n}_{i} := \alpha^{n}_{i}$. 
    \item Define a number $\xi$ as follows : The quantity  $\sum_{i<t} r^{n}_{i}$ increases by $\alpha^{n}_{\lceil2^{n\epsilon}\rceil}$ when $t$ increases by $1$, for values $t>\lceil2^{n\epsilon}\rceil $. To define $\xi$, keep increasing $t$ and stop the first time $\sum_{i<t} r^{n}_{i}>1$ and let $\xi$ be the index before this is seen. I.e., let $\xi$ be such that $\sum_{i\leq \xi} r^{n}_{i}<1 \leq \sum_{i\leq \xi+1} r^{n}_{i}$. Note that $\xi$ is the largest number so that the sum of the $r_{i}$s is at most 1. For $\lceil2^{n\epsilon}\rceil \leq i < \xi$, let $r^{n}_{i} := \alpha^{n}_{\lceil2^{n\epsilon}\rceil}$. Since $  \sum_{i\leq \xi +1} r^{n}_{i} - \sum_{i\leq \xi} r^{n}_{i}= \alpha^{n}_{\lceil2^{n\epsilon}\rceil}$, we have that
\begin{align}
\label{eq:6}
     1-\alpha^{n}_{\lceil2^{n\epsilon}\rceil} \leq \sum_{i\leq \xi} r^{n}_{i} \leq 1.
\end{align}
\item For $\xi \leq i \leq 2^{n}$, let $r^{n}_{i} := 0$. 
\end{enumerate}

So, $r^n_i$ `copies' $\alpha^n_{i}$ from $1 \leq i \leq \lceil2^{n\epsilon}\rceil$, is the constant $\alpha^{n}_{\lceil2^{n\epsilon}\rceil}$ from $ \lceil2^{n\epsilon}\rceil \leq i <\xi$ and is zero at all higher indices.

The following schematic picture describes the function $i \mapsto r^n_i$. The red, green and blue regions depict the definitions in (1),(2) and (3) above respectively\footnote{This schematic picture is not meant to be accurate. E.g., the red region of the graph is meant to depict an arbitrary non-increasing function, not necessarily a  linear non-increasing one.}.  The $y$-axis is the interval $[0,1]$.
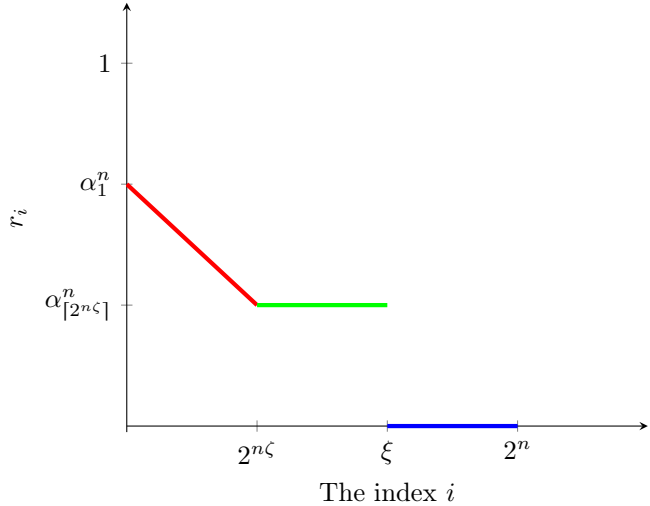
\begin{center}

\begin{tikzpicture}
% Definitions
\tikzmath{
\r1 = 2;
\r2 = 2;
\r3 = 2;
\q1 = 5;
\q2 = 10;
\q3 = 15;
\x1 = \q1; \y1 = 20- \r1*\q1;
\x2 = \q2; \y2 =\r2 * \q2;
\x3 = \q3; \y3 = \r3 * \q3;
 } 
% Axis
\begin{axis}[
axis x line=middle,
axis y line=middle,
ylabel= $r_i$,
xlabel=  The index $i$,
xtick={0,\x1,\x2,\x3},
xticklabels={0,$2^{n\epsilon}$, $\xi$, $2^n$,{ }},
xlabel near ticks,
ytick={0, \y1, \y2, \y3},
yticklabels={0, $\alpha^{n}_{\lceil2^{n\epsilon}\rceil}$, $\alpha^{n}_1$, $1$},
ylabel near ticks,
xmax=\x3+5,
ymax=\y3+5,
xmin=0,
ymin=-0.5
]
% Plots
\addplot[red, ultra thick,domain=0:\q1] {20-\r1*x};
\addplot[green, ultra  thick, domain=\q1:\q2] {10};
\addplot[blue, ultra thick,  domain=\q2:\q3] {0};
\end{axis}
\end{tikzpicture}

\end{center}

The sequence $(r^n_i )_i$ is not a distribution in general as it may not sum to $1$.
In what follows, we will lower bound $H(\rho_n)$ using the Shannon entropy of $(p^n_i )_i$ which is a distribution derived by rescaling $(r^n_i )_i$.

Let $S = \sum_{i\leq 2^{n}} r^{n}_{i} $.
Let $p_{i}^{n}:= S^{-1}r_{i}^{n}$. So, $p=(p_{i}^{n})_{i\leq 2^{n}}$ is a probability distribution on $2^n$. Let $H(p)$ be its Shannon entropy, which we now bound from below. 
 \[H(p) = -\sum_{i< \lceil2^{n\epsilon}\rceil}S^{-1}\alpha^{n}_{i}\text{log}(S^{-1}\alpha^{n}_{i}) - \sum_{ \lceil2^{n\epsilon}\rceil\leq i \leq \xi} S^{-1} \alpha^{n}_{\lceil2^{n\epsilon}\rceil}\text{log}(S^{-1}\alpha^{n}_{\lceil2^{n\epsilon}\rceil}).\]
The first step is to show that the first sum in the above expression is non-negative and thus that the second sum lower bounds $H(p)$ (See \ref{eq:11}).
Recall that we have fixed $n>N_{\delta}$. The definition of $N_\delta$ implies that $\alpha^{n}_{1}\leq \delta$ and hence that for all $i$,
\begin{align}
\label{eq:7}
    \alpha^{n}_{i} \leq \alpha^{n}_{1}\leq \delta<0.5.
\end{align}
Since $\delta<0.5$, 
\begin{align}
\label{eq:8}
    0.5< 1-\delta \leq 1-\alpha^{n}_{\lceil2^{n\epsilon}\rceil} \leq S \leq 1
\end{align}
Putting \ref{eq:6}, \ref{eq:7} and \ref{eq:8} together, for all $i$,
\begin{align}
\label{eq:9}
    \alpha^{n}_{i} \leq \alpha^{n}_{1}<\delta<1-\delta < 1-\alpha^{n}_{\lceil2^{n\epsilon}\rceil} \leq S \leq 1
\end{align}

So, log$(S^{-1}\alpha^{n}_{i})\leq 0$ for all $i$. So, the  first sum in the expression for $H(p)$ is non-negative. This allows us to lower bound $H(p)$ using the second sum in the expression, 
 \[H(p) \geq  - \sum_{ \lceil2^{n\epsilon}\rceil\leq i \leq \xi} S^{-1} \alpha^{n}_{\lceil2^{n\epsilon}\rceil}\text{log}(S^{-1}\alpha^{n}_{\lceil2^{n\epsilon}\rceil}) = -(\xi-\lceil2^{n\epsilon}\rceil)S^{-1} \alpha^{n}_{\lceil2^{n\epsilon}\rceil}\text{log}(S^{-1}\alpha^{n}_{\lceil2^{n\epsilon}\rceil}).\]
 
Since $S^{-1}\geq 1$, 
 
 \begin{align}
 \label{eq:11}
     H(p) \geq  -(\xi-\lceil2^{n\epsilon}\rceil) \alpha^{n}_{\lceil2^{n\epsilon}\rceil}\text{log}(S^{-1}\alpha^{n}_{\lceil2^{n\epsilon}\rceil}). 
 \end{align}

Before moving on with the formal details, we quickly sketch the remaining steps of the proof. We show that the $-(\xi-\lceil2^{n\epsilon}\rceil) \alpha^{n}_{\lceil2^{n\epsilon}\rceil}$ term in the above can be replaced by the constant, $1-2\delta$ (See \ref{eq:10}). We see that $H(\rho_n)\geq H(p)$ by noting that the distribution $(\alpha^n_i)_i$ is more uniform than the distribution $p$. This allows us in \ref{ref} to lower bound $H(\rho_n)$ using the lower bound for $H(p)$ found in \ref{eq:10}. Finally, we use that $\alpha^{n}_{\lceil2^{n\epsilon}\rceil} \leq \delta 2^{-n\epsilon}$ to get the lower bound in \ref{eq:133}. This lower bound gives a contradiction when $n$ tends to infinity. This ends this sketch and we now proceed with the proof.

By the choice of $n>N_{\delta}$, \[S= \sum_{i} r^{n}_{i} = \sum_{i<\lceil2^{n\epsilon}\rceil}\alpha^{n}_{i} + (\xi-\lceil2^{n\epsilon}\rceil)\alpha^{n}_{\lceil2^{n\epsilon}\rceil} \leq \delta + (\xi-\lceil2^{n\epsilon}\rceil)\alpha^{n}_{\lceil2^{n\epsilon}\rceil}.\]
So, $S-\delta \leq (\xi-\lceil2^{n\epsilon}\rceil)\alpha^{n}_{\lceil2^{n\epsilon}\rceil}$.
In conjunction with \ref{eq:9} which says, $1-\delta  \leq S$,
this gives,
\[1-2\delta \leq (\xi-\lceil2^{n\epsilon}\rceil)\alpha^{n}_{\lceil2^{n\epsilon}\rceil}.\]
Since, $-\text{log}(S^{-1}\alpha^{n}_{\lceil2^{n\epsilon}\rceil})\geq 0$, we can put  $1-2\delta$ in place of $ (\xi-\lceil2^{n\epsilon}\rceil)\alpha^{n}_{\lceil2^{n\epsilon}\rceil}$ in \ref{eq:11} to get,
\begin{align}
   \label{eq:10}
   H(p) \geq  -(1-2\delta)\text{log}(S^{-1}\alpha^{n}_{\lceil2^{n\epsilon}\rceil})
\end{align}
I.e.,

\begin{align}
\label{eq:13}
    H(p)\geq  (1-2\delta)[\text{log}(S)-\text{log} (\alpha^{n}_{\lceil2^{n\epsilon}\rceil})].
\end{align}
$\alpha=(\alpha^{n}_{i})_{i}$ and $p=(p^{n}_{i})_{i}$ are both distributions on $2^n$. Note that $r^n_i\geq \alpha^n_i$ whenever $r^n_i > 0$. This is since, $r^n_i= \alpha^n_i$ for $i\leq \lceil 2^{n\epsilon} \rceil$ and $r^n_i = \alpha^n_{\lceil 2^{n\epsilon} \rceil} \geq \alpha^n_i$ for $i\leq \xi$ and $i> \lceil 2^{n\epsilon} \rceil$ and $r^n_i = 0$ for $i>\xi$. This is depicted in the schematic picture where the graph of $(\alpha^n_i)_i$ (in dotted black) is superimposed on the graph of $(r^n_i)_i$ from earlier. The graph of $(\alpha^n_i)_i$ is meant to schematically depict a non-increasing function. The graph of $(r^n_i)_i$ (the red part) agrees with that of $(\alpha^n_i)_i$ for indices $i\leq \lceil 2^{n\epsilon} \rceil$ , the graph of $(r^n_i)_i$ (the green part) dominates that of $(\alpha^n_i)_i$ for $ \lceil 2^{n\epsilon} \rceil \leq i <\xi$ while 
the graph of $(\alpha^n_i)_i$ dominates that of $(r^n_i)_i$ (the blue part) for $i \geq \xi$.

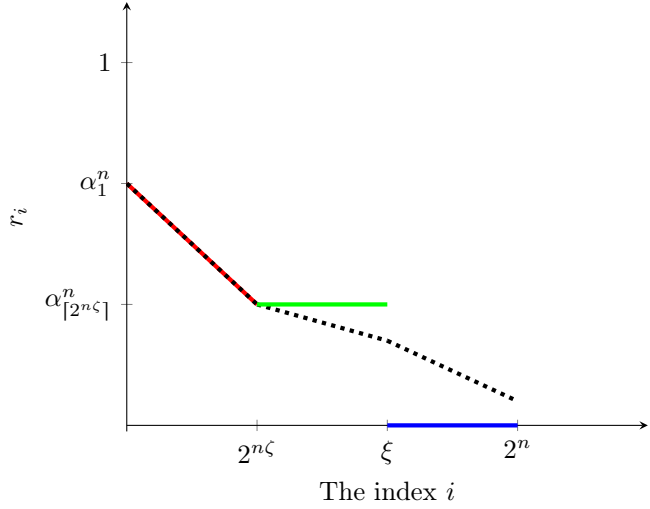
\begin{center}

\begin{tikzpicture}
% Definitions
\tikzmath{
\r1 = 2;
\r2 = 2;
\r3 = 2;
\q1 = 5;
\q2 = 10;
\q3 = 15;
\x1 = \q1; \y1 = 20- \r1*\q1;
\x2 = \q2; \y2 =\r2 * \q2;
\x3 = \q3; \y3 = \r3 * \q3;
 } 
% Axis
\begin{axis}[
axis x line=middle,
axis y line=middle,
ylabel= $r_i$,
xlabel=  The index $i$,
xtick={0,\x1,\x2,\x3},
xticklabels={0,$2^{n\epsilon}$, $\xi$, $2^n$,{ }},
xlabel near ticks,
ytick={0, \y1, \y2, \y3},
yticklabels={0, $\alpha^{n}_{\lceil2^{n\epsilon}\rceil}$, $\alpha^{n}_1$, $1$},
ylabel near ticks,
xmax=\x3+5,
ymax=\y3+5,
xmin=0,
ymin=-0.5
]
% Plots
\addplot[red, ultra thick, domain=0:\q1] {20-\r1*x};
\addplot[black, ultra thick, dotted, domain=0:\q1] {20-\r1*x};
\addplot[black, ultra  thick, dotted, domain=\q1:\q2] {13-\r1*x*0.3};
\addplot[green, ultra  thick,  domain=\q1:\q2] {10};
\addplot[blue, ultra thick, domain=\q2:\q3] {0};
\addplot[black, ultra  thick, dotted, domain=\q2:\q3] {17 -\r1*x*0.5};
\end{axis}
\end{tikzpicture}

\end{center}

Let $i$ be such that $p^n_{i}>0$. Then $r^n_i >0$ too and we have that $ p^n_{i} = S^{-1}r^n_{i}\geq r^n_{i} \geq \alpha^n_{i}$. So, the distribution $p$ is greater than or equal to $\alpha$ at points $i$ such that $p^n_i >0$. It follows from the definition of a distribution that $\alpha$ is greater than or equal to $p$
at points $i$ such that $p^n_i = 0$. This means that $\alpha$ is more uniform than $p$. A result from the book by Cover and Thomas (exercise 2.28, pg.50 in \cite{cover2012elements}) thus implies that $H(\alpha) \geq H(p)$. So, by \ref{eq:13}, we have,
\begin{align}
\label{ref}
    H(\rho_{n})= H(\alpha) \geq   (1-2\delta)[\text{log}(S)-\text{log}(\alpha^{n}_{\lceil2^{n\epsilon}\rceil})]
\end{align}

Further, note that $\alpha^{n}_{\lceil2^{n\epsilon}\rceil} \leq \delta 2^{-n\epsilon}$.\footnote{If not, then, for all  $i \leq \lceil2^{n\epsilon}\rceil$, $\alpha^{n}_{i} \geq \alpha^{n}_{\lceil2^{n\epsilon}\rceil} > \delta 2^{-n\epsilon}$. This would give 
$ \sum_{i \leq \lceil2^{n\epsilon}\rceil}\alpha^{n}_{i}>\lceil2^{-n\epsilon}\rceil2^{-n\epsilon} \delta \geq \delta,$ contradicting that $n>N_{\delta}$.} So, taking log on both sides:
$\text{log} (\alpha^{n}_{\lceil2^{n\epsilon}\rceil}) \leq \text{log} (\delta) + \text{log} (2^{-n\epsilon})$ and thus that:
\begin{align}
\label{eq:12}
    -\text{log} (\alpha^{n}_{\lceil2^{n\epsilon}\rceil}) \geq -\text{log} (\delta) +  n\epsilon.
\end{align}

Recalling that $1-\delta<S\leq 1$ we have that log$(S)>$log$(1-\delta)$. By this, \ref{ref} and \ref{eq:12}, 
\begin{align}
\label{eq:133}
    H(\rho_{n}) >   (1-2\delta)[\text{log}(1-\delta)-\text{log}(\alpha^{n}_{\lceil2^{n\epsilon}\rceil})]\geq (1-2\delta)[\text{log}(1-\delta)-\text{log}(\delta) +  n\epsilon]
\end{align}
So, 
\[\dfrac{H(\rho_{n})}{n} > (1-2\delta)[\dfrac{ \text{log}(1-\delta)-\text{log}(\delta) +  n\epsilon}{n}].\]
As the $n>N_{\delta}$ here was arbitrary, the above holds for all $n>N_{\delta}$ and we can let $n$ go to infinity.
Using that  $\liminf(x_n + y_n) \geq \liminf (x_n) + \liminf(y_n)$  we get:
\[ H(\rho) = \liminf_{n}\dfrac{H(\rho_{n})}{n} > (1-2\delta)[\liminf_{n}\dfrac{\text{log}(1-\delta)-\text{log} (\delta)}{n} + \epsilon]=(1-2\delta)\epsilon.\]
As $\delta<0.5$ was arbitrary, we have that $H(\rho)>(1-2\delta)\epsilon$ for all $\delta<0.5$. By assumption we have that $H(\rho)<\epsilon$. Let $\nu>0$ be such that $ \epsilon - \nu = H(\rho)$.

Fix a sufficiently small $\delta_0 <0.5$ such that $(1-2\delta_0)\epsilon> \epsilon - \nu$. Then, we have that $H(\rho)>(1-2\delta_0)\epsilon> \epsilon - \nu = H(\rho)$. This gives the desired contradiction. 
\end{proof}

\section{Entropy rate and quantum s-tests}
\label{rqr}
Let $s\in [0,1]$. A Solovay s-test $S$ is a computably enumerable subset of $2^{<\omega}$ such that $\sum_{\sigma \in S} 2^{-s|\sigma|} < \infty$. A $X\in 2^\omega$ is said to be covered by $S$ if $X\in \llbracket \sigma \rrbracket$ for infinitely many $\sigma$'s in $S$ (See Definition 13.5.6 in \cite{misc1}). The limiting `Kolmogorov complexity rate' of bitstrings is closely related to Solovay s-tests:  
$\liminf_n K(X\upharpoonright n)/n = \liminf_n C(X\upharpoonright n)/n = \inf \{ s: X $ is covered by a Solovay s-test$\}$ (See pg.603 and Thm.13.3.4 in \cite{misc1}).

We extend the notion of a Solovay s-test to the quantum realm and  relate this extension to the entropy rate of states. This extension will need a slight reframing of the definition of a Solovay s-test.

Let $S$ be a Solovay s-test. For each $n\geq 0$, let $S^n := S \cap 2^n$. We have
\[\infty> \sum_{\sigma \in S} 2^{-s|\sigma|} = \sum_n \sum_{\sigma \in S^n} 2^{-sn}=\sum_n 2^{-sn}|S^n|.\]
If $X$ is covered by $S$, then it must be the case that $X\in \llbracket S_n \rrbracket$ for infinitely many $S_n$. We hence have the equivalent definitions:
\begin{defn}
\label{sol} Let $s\in [0,1]$.
A Solovay s-test is a
uniformly computably enumerable sequence, $S=(S^n)_n$  such that $S^n \subseteq 2^n$ and $\sum_{n} 2^{-sn}|S^n| < \infty$. A $X\in 2^\omega$ is said to be covered by $S$ if $X\in \llbracket S^n \rrbracket$ for infinitely many $n$.

\end{defn}

The quantum analogue of  $S^n$ is a special projection $T^n$ on $\mathbb{C}^{2^n}$ such that $\tr(T^n) = |S^n|$ (See Defn.3.2 in \cite{unpublished} and Section \ref{qardefn}). Also, the quantum analogue of $X\in \llbracket S^n \rrbracket$ is that $\rho(T^n)>\delta$ for some $\delta>0$ independent of $n$.
Generalizing Definition \ref{sol} to the quantum setting hence yields the following notion:
\begin{defn}
    Let $s \in [0,1]$. A  quantum s-test (q s-test) is a computable sequence of special projections $(T^m)_{m=1}^\infty$ where $T^m$ is on $\mathbb{C}^{2^{n_m}}$ and such that $\sum_m 2^{-s n_m}\tr(T^m) < \infty $. A state is said to be covered by $(T^m)_m$ if there is a $\delta>0$ such that $\rho(T^m)>\delta$ for infinitely many $m$.
\end{defn}

\begin{thm}
  Let $\rho = (\rho_n)_n$ be computable and let $s\in [0,1]$. If $s>H(\rho)$ then $\rho$ is covered by a q s-test. In particular, $H(\rho)\geq \inf \{s: \rho$ is covered by a q s-test $\}$.   
\end{thm}
\begin{proof}
Let $\rho=(\rho_n)_n$ be a computable state and let
\[\rho_{n} = \sum_{i\leq 2^n} \alpha^{n}_{i}|\psi^{n}_{i}\big> \big< \psi^{n}_{i}|.
\]
Let the eigenvalues be labelled such that $\alpha^{n}_{1} \geq \alpha^{n}_{2}\dots \geq\alpha_{2^{n}}^{n}$.

Fix a rational $t$ such that $H(\rho)<t<s$. By Lemma \ref{lemmagene}, there is a $\delta>0$ such that for infinitely many $n$, \[\sum_{i \leq \lceil2^{n t}\rceil}\alpha^{n}_{i} > \delta.\] 
    We build a q s-test $(S^m)_m$. We describe the construction of a special projection $S_m$ uniformly in $m$.

\emph{Construction of $S^m$:}
Find a $n$ such that both of the following hold:
\begin{enumerate}

    \item $\sum_{i \leq \lceil2^{n t}\rceil}\alpha^{n}_{i} > \delta$.
    
    \item $\dfrac{2^{n t}+1}{2^{ns}}<2^{-m}$. 
\end{enumerate}    
Set
\[ S^{m} := \sum_{i\leq \lceil2^{n t}\rceil} |\psi^{n}_{i}\big> \big< \psi^{n}_{i}|. \] 
Denote the $n$ found above by $n_m$.

\emph{Verification:}

By Lemma \ref{lemma}, there are infinitely many $n$ such that condition (1) holds. Further, as $\rho$ is computable, the satisfaction of condition (1) for a given $n$ can be computably checked. Condition (2) holds for almost every $n$ since $t<s$. So $n_m$ can be computed uniformly in $m$. Note that $\tr(S^m) \leq 2^{n t}+1$. Since $n_m$ satisfies the second condition, $2^{-n_m s}\tr(S^{m})<2^{-m}$ thus showing that $\sum_m 2^{-n_m s}\tr(S^m) < \infty$. As $n_m$ satisfies the first condition, $\rho(S^m)=\tr(\rho_{n_m}S^m)>\delta$ for all $m$ thus showing that $(S^m)_m$ covers $\rho$. 

\end{proof}

We leave it open whether $H(\rho)=\inf \{s: \rho $ is covered by a q s-test $\}$ for computable states. If true, this would mirror the relation between the initial segment Kolmogorov complexity and Solovay s-tests mentioned earlier. Lemma \ref{power} shows this equality for a certain special class of states defined in Definition \ref{defnd}. This uses the quantum typical subspace theorem, a key result in quantum channel coding: 
\begin{thm}
\label{tst}
    Let $D$ be a density matrix on $\C^{t}$ and let $r<H(D)$. If $(S(n))_n$ is any sequence of projectors such that $S(n)$ is on $\C^{t^n}$ and $\tr(S(n))\leq 2^{nr}$ then $\lim_n \tr(S(n) \bigotimes_{i=1}^n D )=0 $ (See (3) in Theorem 12.5 in Section 12.2.2 in \cite{nielsen2010quantum}).
\end{thm}

For any natural number $a$ let $PT_a$ denote the operator given by applying the $PT_{\mathbb{C}^2}$ operator (See Definition.\ref{state}) $a$ times. I.e. $PT_a$ is the partial trace operator which traces out the last $a$ many qubits of a density matrix. 
\begin{defn}
\label{defnd}
   
Fix a density matrix $d$ on $\C^{2^k}$ and let $\rho^d= (\rho^d_z)_z$ be the state given by $\rho^d_{nk} := \bigotimes_{i=1}^n d$ for all $n$. For $z$ which is not a multiple of $k$, let $\rho_z := \text{PT}_{nk-z}(\rho_{nk})$ where $nk$ is the least multiple of $k$ greater than $z$.     
\end{defn}
We are now ready to state the result:
\begin{lem}
\label{power}
Fix a density matrix $d$ on $\C^{2^k}$.
If $s<H(\rho^d)$ then $\rho^d$ cannot be covered by a q s-test. In particular,  $H(\rho^d)\leq \inf \{s: \rho $ is covered by a q s-test $\}$
\end{lem}

\begin{proof}
We denote $\rho^d$ by $\rho$ to simplify notation.
As $H(\rho_{nk})=H(\bigotimes_{i=1}^{n} d ) = nH(d)$, we have that 
\[\lim_n \dfrac{H(\rho_{nk})}{nk}= \dfrac{H(d)}{k}.\]
Since $(H(\rho_{nk})/nk)_{n=1}^\infty$ is a subsequence of $(H(\rho_z)/z)_{z=1}^\infty$ we get that $H(\rho)\leq H(d)/k$ by properties of the lim inf. So $s<H(d)/k$ and hence $sk<H(d)$.

Suppose towards a contradiction that $\rho$ is covered by a q s-test $(T^m)_m$ where each $T^m$ is on $\mathbb{C}^{2^{u_m}}$. We may assume that each $u_m$ is a multiple of $k$.\footnote{Suppose that $nk\geq u_m$ is the smallest multiple of $k$ greater than $u_m$. Let $C^m$ be defined as $C^m:= T^m \otimes \bigotimes_{i=1}^{nk-u_m} I$. Note that $\tau(T^m)=\tau(C^m)$ and that $\rho(T^m)=\rho(C^m)$. This shows that replacing $T^m$ by $C^m$ for all $m$ gives a q s-test that covers $\rho$.} Let $u_m = n_m k$ for all $m$.

By omitting finitely many $T^m$'s, we may assume that $\sum_m 2^{-s u_m}\tr(T^m)<1$.

So for all $m$ we have that $\tr(T^m)< 2^{s u_m}= 2^{n_m(sk)}$ and $T^m$ is on $\mathbb{C}^{2^{k n_m}}=\mathbb{C}^{(2^{k})^{n_m}}$.

Putting $D=d, t=2^k, n=n_m, S(n)=T^m, r=sk$ in Theorem \ref{tst} and recalling that $sk<H(d)$ gives that $\lim_m \tr(T^m \bigotimes_{i=1}^{n_m}d)=0$.

Noting that $ \bigotimes_{i=1}^{n_m}d = \rho_{n_m k}$ we get that $\lim_m \tr(T^m \rho_{n_m k})=\lim_m \rho(T^m)=0$. As $(T^m)_m$ covers $\rho$ there is a $\delta>0$ such that $\rho(T^m)>\delta$ for infinitely many $m$. This is a contradiction.
\end{proof}

\section{The singular value decomposition}
\label{svd}
The singular value decomposition (SVD) is an important linear algebraic notion.
We use the following consequence of the singular value decomposition(See Theorem 3.1 in the textbook by Blum, Hopcroft and Kannan \cite{Blum_Hopcroft_Kannan_2020}).
\begin{thm}
\label{thm:bhk}
Let $A$ be a $n \times q$ complex valued matrix with singular vectors $v_1, v_2, \dots v_r$ and corresponding singular values $\sigma_1 \geq \sigma_2, \dots \geq \sigma_r$. Let $k \leq r$ and $w_1,w_2, \dots, w_k$ be any orthonormal set. Then, 
\[ \sum_{i \leq k} |Av_{i}|^{2} \geq \sum_{i \leq k} |Aw_{i}|^{2}.
\]
\end{thm}
Sections \ref{ls} and \ref{ui}  will frequently use the following result which is implied by Theorem \ref{thm:bhk}.
\begin{thm}
\label{thm:10}
    Let $G$ be a rank $k$ Hermitian projection and let $d$ be a density matrix on $\mathbb{C}^{2^n}$. Then $\tr(G d)$ is bounded above by the sum of the first $k$ largest eigenvalues of $d$.
\end{thm}
\begin{proof}
Let
\[ d  = \sum_{i\leq 2^n} \alpha_{i} |\psi_i\big> \big< \psi_i|,\]
be the eigendecomposition of $d$ (See Definition \ref{dens}). Suppose that the eigenvalues are labelled such that $\alpha_1 \geq \cdots \geq \alpha_{2^n}$.
Define
\[ \sqrt d  = \sum_{i\leq 2^n} \sqrt \alpha_{i} |\psi_i\big> \big< \psi_i|,\]
which is well defined since $\alpha_i \geq 0$ for all $i$.

Note that $\sqrt d$ has eigenpairs ($\psi_{i},\sqrt \alpha_{i}$) and is Hermitian with non-negative eigenvalues. So its eigenvector-eigenvalue pairs are identical to its singular vector-singular value pairs (See page 30 in \cite{Blum_Hopcroft_Kannan_2020}).

Let $k \leq 2^n$ and let $G$ be an arbitrary Hermitian projection of rank $k$ on $\mathbb{C}^{2^n}$. So there is an orthonormal set $w_1,w_2, \cdots, w_k$ in $\mathbb{C}^{2^n}$ such that  $G=\sum_{i \leq k} |w_{i}\big>\big<w_{i}| $ is the Hermitian projection onto the subspace spanned by $w_1,w_2, \dots, w_k$. Since the first $k$ singular vectors of $\sqrt d$ (arranged in decreasing order of their corresponding singular values) are $\psi_1, \dots, \psi_k$, Theorem \ref{thm:bhk} gives that:

\begin{align}
\label{eq:14}
    \sum_{i \leq k} |\sqrt d \psi_{i}|^{2} \geq \sum_{i \leq k} |\sqrt d w_{i}|^{2}
\end{align}
 
Since $\sqrt d$ is Hermitian, $\big<x|\sqrt d y \big> = \big< (\sqrt d)^{*} x| y \big> = \big< \sqrt d  x| y \big>$. So

\[ \tr( d G)= \sum_{i \leq k}   \tr(  d |w_{i}\big>\big<w_{i}|) = \sum_{i \leq k}   \big<w_{i}|  d |w_{i}\big> = \sum_{i \leq k}   \big<\sqrt d w_{i}|  \sqrt d w_{i}\big> = \sum_{i \leq k} |\sqrt d w_{i}|^{2}.
\]
So, by inequality \ref{eq:14},
\begin{align}
\label{eq:15}
     \tr(dG)=  \sum_{i \leq k} |\sqrt d w_{i}|^{2}\leq \sum_{i \leq k} |\sqrt d \psi_{i}|^{2}  = \sum_{i \leq k}   \alpha_{i}.
\end{align}

\end{proof}

\section{Initial segment entropy and strong quantum randomness.}
\label{ls}
In contrast to the arguments made so far which relied on the computability of the state, the results of this section hold for any state, regardless of its computational hardness.
As is standard in computability, quantum Schnorr randomness can be relativized to any oracle. For any $X\in 2^\omega$, a $X$-q-S test will denote a q-S test in which the computability is with respect to the oracle $X$. A state is said to be $X$-quantum Schnorr random if it passes all $X$-q-S tests at all $\delta >0$.

We define strong quantum randomness, a randomness notion which implies $X$-quantum Schnorr randomness for any $X$.

Theorem \ref{thm:9} gives a condition on the initial segment von Neumann entropy which implies strong quantum randomness.

\begin{defn}
\label{weakqt}
A quantum null condition is a sequence  $(T^{m})_{m=1}^\infty$ of special projections such that $\lim_m \tau(T^m)=0$.

A state
$\rho=(\rho_n)_{n=1}^\infty$ is said to satisfy a quantum null condition $(T^{m})_{m=1}^\infty$ if $\inf \{\rho(T^m): m\} = 0$. A state $\rho$ is defined to be strong quantum random if it satisfies all quantum null conditions. I.e., $\rho$ is strong quantum random if for all quantum null conditions $(T^m)_m$ for any $\delta>0$ there is a $m$ such that $\rho(T^m)\leq \delta$. 
\end{defn}
\begin{remark}
\label{sqr}
Note that strong quantum randomness implies $X$-quantum Schnorr randomness for any $X\in 2^\omega$: Suppose that $\rho$ is strong quantum random. For a contradiction, suppose that $(G^m)_m$ is a $X$-q-S test which $\rho$ fails at $\delta$. There is hence a subsequence $(G^{k_m})_m$ such that $\rho(G^{k_m})>\delta$ for every $m$. Since $\sum_m \tau(G^{k_m})$ is finite we have that $\lim_m \tau(G^{k_m})=0$ and hence that $(G^{k_m})_m$ is a quantum null condition. This gives a contradiction since $\inf\{\rho(G^{k_m}):m\}=\delta>0$.
\end{remark}

Note that the tracial state (See \ref{def:tr}) is strong quantum random by definition.
This condition in Theorem \ref{thm:9} says that there is an absolute constant $c$, such that the von Neumann entropy of the length $n$ initial segment infinitely often comes `$c$ close' to $n$. Recall that the von Neumann entropy of the length $n$ initial segment of the tracial state is $n$. So, if $\rho$ satisfies the condition, then it is infinitely often `equal' to the tracial state, upto a `finite difference' given by $c$. It is thus quite natural to expect that this condition implies strong quantum randomness. 

\begin{thm}
\label{thm:9} Let $\rho$ be any state. If there exists a natural number $c$ such that $H(\rho_{n})>n-c$ for infinitely many $n$ then $\rho$ is strong quantum random.
\end{thm}
\begin{proof}

Let $\rho=(\rho_{n})_{n}$ be a state and let $c>0$ be a natural number such that $H(\rho_n)>n-c$ for infinitely many $n$. Each $\rho_n$ can be written as \[\rho_{n} = \sum_{i\leq 2^n} \alpha^{n}_{i}|\psi^{n}_{i}\big> \big< \psi^{n}_{i}|,
\]
as in Definition \ref{dens}.
We assume that the eigenvalues are labelled such that $\alpha^{n}_{1} \geq \alpha^{n}_{2}\dots \geq\alpha_{2^{n}}^{n}$.

Suppose for a contradiction that $\rho$ does not satisfy a quantum null condition $(Z^m)_m$. I.e., there is a $\delta>0$ such that $\rho(Z^m)>\delta$ for all $m$. Since $\lim_m \tau(Z^m)=0$, pick a subsequence, $(T^m)_m$ of the $(Z^m)_m$ sequence such that $\tau(T^m) < 2^{-m}$ for all $m$.

For all $m$ let $N_m$ be such that $T^m$ is a special projection on $\mathbb{C}^{2^{N_m}}$.

For each $m$ let $(G^m_n)_{n=N_m}^\infty$ be a sequence of special projections such that $G^m_n$ is on $\mathbb{C}^{2^n}$ and
\[G^m_n = T^m \otimes I_{n-N_m}.\]
Since $\tau(T^m)=2^{-N_m}\tr(T^m)<2^{-m}$, we get that $\tr(T^m)<2^{N_m - m}$. So for any $n$,
\begin{align}
    \label{tr}
    \tr(G^m_n) = \tr(T^m)\tr(I_{n-N_m}) < 2^{N_m - m}2^{-N_m}=2^{n-m}.
\end{align}

By the definition of the partial trace and of a state (See Definition \ref{state}),
\begin{align}
    \label{del}
    \tr(\rho_n G^m_n)=\tr(\rho_n ( T^m \otimes I_{n-N_m}))=\tr(\rho_{N_m} T^m)>\delta.
\end{align}

For all $m$ and $n\geq N_m$ let,

\[S_{m,n} = \sum_{i \leq 2^{n-m}}   \alpha^{n}_{i},\]
be the sum of the first $2^{n-m}$ many largest eigenvalues of $\rho_n$.

\emph{Proof sketch:}

For all $m$ and all $n\geq N_m$, we have that rank$(G^m_n)=\tr(G^m_n)<2^{n-m}$ by \ref{tr}. 
Putting $d=\rho_n$ and $k=2^{n-m}$ in Theorem \ref{thm:10} gives that $\tr(G^m_n \rho_n)\leq S_{m,n}$  for any $m$ and $n\geq N_m$. Since $\tr(G^m_n \rho_n)>\delta$ for all $m$ and $n\geq N_m$ (See \ref{del}) we get that $S_{m,n} > \delta$ for all $m$ and all $n\geq N_m$.

For all $m$ and all $n\geq N_m$, we construct a piecewise uniform distribution, $(r^m_n(i))_i$, on $\{1,\cdots, 2^{n}\}$ given by averaging, $(\alpha^n_i)_i$ (the distribution given by the eigenvalues of $\rho_n$) over the block $\{1,\cdots, 2^{n-m}\}$ and over the block $\{2^{n-m}+1,\cdots,2^n\}$. The distribution $(r^m_n(i))_i$ is uniform on each block and is thus more uniform than $(\alpha^n_i)_i$. This implies that $H((r^m_n(i))_i)$ is an upper bound for $H(\rho_n )$. Using this and that $S_{m,n} > \delta$, we get that, $H(\rho_n ) \leq  1- mS_{m,n} +n < 1- m\delta +n$  for all $m$ and all $n\geq N_m$. We get a contradiction by letting $m$ tend to infinity.

\emph{Proof details:} Fix $m$ and $n>N_m$. Theorem \ref{thm:10} and inequality \ref{tr} implies that $\tr(\rho_{n} G^{m}_{n})\leq S_{m,n}$. Inequality \ref{del} gives
\begin{align}
\label{eq:16}
    \delta< \tr(\rho_{n} G^{m}_{n})\leq S_{m,n}.
\end{align} 

Let $(r_{n}^{m}(i))_{i\leq 2^n}$ be the distribution on $\{1,2,....2^n\}$ defined as:

\begin{enumerate}
    \item $r^m_n (i):= S_{m,n} 2^{m-n}$ if $i\leq 2^{n-m}$
    \item $r^m_n (i)= (1-S_{m,n})/2^{n}(1-2^{m})$ if $ 2^{n-m}<i\leq 2^n$.
\end{enumerate}
 
In short, the first piece of $(r_{n}^{m}(i))_{i\leq 2^n}$ is obtained by averaging $(\alpha^n_i)_{i\leq 2^n}$ over $\{1,\cdots,2^{n-m}\}$ while the second piece of $(r_{n}^{m}(i))_{i\leq 2^n}$ is obtained by averaging $(\alpha^n_i)_{i\leq 2^n}$ over $\{2^{n-m}+1,\cdots, 2^n\}$. 
 
So, $(r_{n}^{m}(i))_{i\leq 2^n}$ is uniform on each of the blocks, $\{1,...,2^{n-m}\}$ and $\{ 2^{n-m}+1,....,2^{n}\}$. Its total probability mass on the first block is $S_{m,n}$ and on the second is $1-S_{m,n}$.

The distribution $(\alpha^n_i)_{i\leq 2^n}$ also has a probability mass of $S_{m,n}$ on the first block and a mass of $1-S_{m,n}$ on the second block. 

Both distributions $(r_{n}^{m}(i))_{i\leq 2^n}$ and $(\alpha^n_i)_{i\leq 2^n}$ hence assign equal probability mass to the blocks. The distributions however, differ within each block: $(r_{n}^{m}(i))_{i\leq 2^n}$ is more uniform than $(\alpha^n_i)_{i\leq 2^n}$ \emph{within} each block.

Another way to picture this is to note that $(r_{n}^{m}(i))_{i\leq 2^n}$ is obtained by uniformly distributing $(\alpha^n_i)_{i\leq 2^n}$'s total mass of $S_{m,n}$ over $\{1,\cdots,2^{n-m}\}$ and by uniformly distributing $(\alpha^n_i)_{i\leq 2^n}$'s total mass of $1-S_{m,n}$ over $\{ 2^{n-m}+1,\cdots,2^{n}\}$.

The picture schematically depicts this. The dashed line is the graph of $(\alpha_{n}(i))_{i}$ and is meant to depict a decreasing function. The red and blue lines are the values of $r^m_n(i)$ on the two blocks $\{1,\cdots,2^{n-m}\}$ and $\{ 2^{n-m}+1,\cdots,2^{n}\}$ respectively. The height of the red line is $S_{m,n} 2^{m-n}$ and that of the blue line is $(1-S_{m,n})/(1-2^{n-m})$. The red line is the average of the values of the black dashed line over the block $\{1,\cdots,2^{n-m}\}$ while the blue line is the average of the values of the black dashed line over the block $\{2^{n-m}+1,\cdots, 2^n\}$.

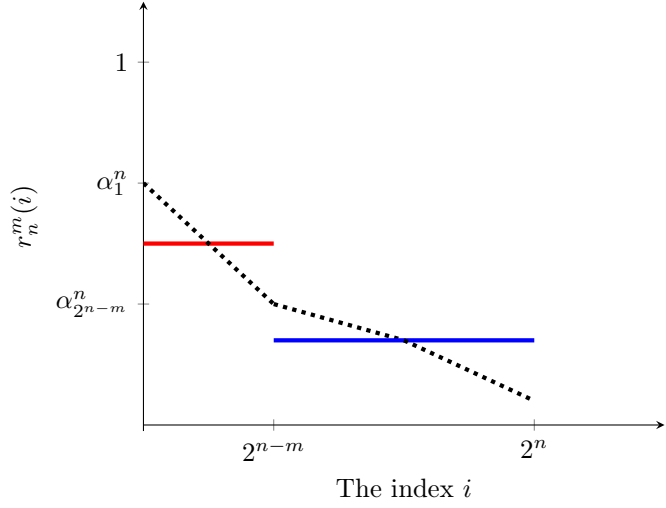
\begin{center}

\begin{tikzpicture}
% Definitions
\tikzmath{
\r1 = 2;
\r2 = 2;
\r3 = 2;
\q1 = 5;
\q2 = 10;
\q3 = 15;
\x1 = \q1; \y1 = 20- \r1*\q1;
\x2 = \q2; \y2 =\r2 * \q2;
\x3 = \q3; \y3 = \r3 * \q3;
 } 
% Axis
\begin{axis}[
axis x line=middle,
axis y line=middle,
ylabel= $r^m_n(i)$,
xlabel=  The index $i$,
xtick={0,\x1, \x3},
xticklabels={0,$2^{n-m}$, $2^n$, $2^n$,{ }},
xlabel near ticks,
ytick={0, \y1, \y2, \y3},
yticklabels={0, $\alpha^{n}_{2^{n-m}}$, $\alpha^{n}_1$, $1$},
ylabel near ticks,
xmax=\x3+5,
ymax=\y3+5,
xmin=0,
ymin=-0.5
]
% Plots
\addplot[red, ultra thick, domain=0:\q1] {15};
\addplot[black, ultra thick, dotted, domain=0:\q1] {20-\r1*x};
\addplot[black, ultra  thick, dotted, domain=\q1:\q2] {13-\r1*x*0.3};
\addplot[blue, ultra  thick,  domain=\q1:\q2] {7};
\addplot[blue, ultra thick, domain=\q2:\q3] {7};
\addplot[black, ultra  thick, dotted, domain=\q2:\q3] {17 -\r1*x*0.5};
\end{axis}
\end{tikzpicture}

\end{center}

 So, by exercise 2.28 on page 50 of the book \cite{cover2012elements} we have that
 \begin{align}
 \label{eq:17}
     H(\rho_n)\leq H((r_{n}^{m}(i))_{i\leq 2^n})
 \end{align}
 Recall that this holds for $\forall m$ and $\forall n >N_{m}$.

 Let the distribution $(r_{n}^{m}(i))_{i\leq 2^{n}} $ be denoted by $r^{m}_{n}$.
 We now bound $H(r^{m}_{n})$ from above to get a contradiction.
 
 \[H(r^{m}_{n})=-\bigg[ 2^{n-m}(S_{m,n} 2^{m-n}\text{log}(S_{m,n} 2^{m-n}))+(2^{n}-2^{n-m})\dfrac{(1-S_{m,n})}{2^{n}(1-2^{m})}\text{log}(\dfrac{(1-S_{m,n})}{2^{n}(1-2^{m})})\bigg]\]
 
 This simplifies to: 
 \[H(r^{m}_{n})=h(S_{m,n}) - mS_{m,n} + n +(1-S_{m,n})\text{log} (1-2^{-m})\]
 where $ h(S_{m,n}) := -S_{m,n} \text{log}(S_{m,n})-(1-S_{m,n})\text{log} (1-S_{m,n})$ is a real number in $[0,1]$. Since
$\text{log}(1-2^{-m})<0$ and $(1-S_{m,n})>0$, we have $(1-S_{m,n})\text{log} (1-2^{-m})<0$. So, we can replace it by $0$ to get an upper bound, 
\begin{align}\label{eq:18}
    H(r^{m}_{n}) \leq 1 - m S_{m,n} + n.
\end{align}
By \ref{eq:17} and \ref{eq:18}, we get that for all $m$ for all $n>N_{m}$, 
\begin{align}
\label{eq:19}
    H(\rho_{n}) \leq 1 - mS_{m,n} + n
\end{align}
Recall that by assumption, we have that:

\begin{align}
\label{eq:220}
    \exists^{\infty}n ,   H(\rho_n) \geq n- c.
\end{align}

Now, for all $m$, \ref{eq:19} holds for almost every $n$ and \ref{eq:220} holds for infinitely many $n$. So, for all $m$, both \ref{eq:19} and \ref{eq:220} simultaneously hold for infinitely many $n$. I.e, \[\forall m \exists^{\infty} n,\text{ such that } n-c \leq  H(\rho_{n})\leq 1 - mS_{m,n} + n\]
So we get that
$\forall m \exists^{\infty} n,  -c \leq    1 - mS_{m,n}$, and hence that
\[\forall m \exists^{\infty} n,\text{ such that }   c \geq    -1 + mS_{m,n}.  \]
Noting that $S_{m,n}>\delta$ for $n>N_{m}$, we get that
\[\forall m \exists^{\infty} n,\text{ such that }   c \geq    -1 + mS_{m,n} \geq -1 + m \delta .  \]
So,
\[\forall m,    c +1 \geq   m \delta, \]
which is a contradiction.
\end{proof}

\subsection{Examples of states satisfying the initial segment condition}
\label{not}

Recall that $2^\omega$ can be identified with the interval $[0,1)$ (See 1.8.10 in \cite{misc}). For a string $\sigma \in 2^{<\omega}$ let $[\sigma] \subseteq [0,1)$ denote the image of  $\llbracket \sigma \rrbracket \subset 2^\omega$ under this identification.

For a continuous function $g$ on $(0,1)$, Reimann Sum$[g, T $ Mesh Size $= \nu]$ denotes the Reimann sum of $g$ computed using Mesh size $\nu$ and the sample points in $T$ (See pg.164, Chapter 3, Section 2 in \cite{PughCharlesChapman2015RMA}).  
Any state differing from the tracial state, $\tau$ (Recall Definition \ref{def:tr}), at only finitely many qubits clearly satisfies the hypothesis of Theorem \ref{thm:9}. We construct more examples of states satisfying the hypothesis of Theorem \ref{thm:9}: Let $f$ be any continuous function on $(0,1)$ satisfying

\[\int_{0}^{1}f(s)ds =1 \text{ and } -\infty < -\int_{0}^{1}f(s)\text{log}(f(s))ds < \infty.\]

For example, let\[f(x)= \dfrac{2}{x(1-\oln x)^{3}},\] on $(0,1)$ where $\oln$ stands for the natural logarithm.
Define a diagonal state $\rho = (\rho_{n})_{n}$ as follows. Fix $n$. For all $\sigma \in 2^n$, let
\[\alpha_{\sigma} = \int_{[\sigma]} f(s)ds,\] and let
\[\rho_{n}= \sum_{\sigma \in 2^{n}} \alpha_{\sigma} |\sigma\big>\big<\sigma|.\]
$\rho=(\rho_n)_n$ is a state since $\alpha_{\sigma}= \alpha_{\sigma 1} + \alpha_{\sigma 0}$ by definition and since $\int_{0}^{1}f(s)ds =1 $.  Now we show that $\rho$ satisfies the hypothesis of \ref{thm:9}. For any $n$, by definition of the $\alpha$s, we have, \[H(\rho_{n}) = -\sum_{\sigma \in 2^{n}}\int_{[\sigma]} f(s)ds \text{log}(\int_{[\sigma]} f(s)ds)\]
By the mean-value theorem and continuity of $f$, for all $\sigma$ there is a $x_{\sigma}\in [\sigma]$ such that
\[\int_{[\sigma]} f(s)ds = 2^{-n}f(x_{\sigma})\]
So,
\[H(\rho_{n}) = -\sum_{\sigma \in 2^{n}}2^{-n}f(x_{\sigma}) \text{log}(2^{-n}f(x_{\sigma})) \]
\[=-\sum_{\sigma \in 2^{n}}2^{-n}f(x_{\sigma}) (-n+\text{log}(f(x_{\sigma}))) \]

\[=-\sum_{\sigma \in 2^{n}}2^{-n}f(x_{\sigma}) \text{log}(f(x_{\sigma})) + n \sum_{\sigma \in 2^{n}}2^{-n}f(x_{\sigma}) \]

\[=-\sum_{\sigma \in 2^{n}}2^{-n}f(x_{\sigma}) \text{log}(f(x_{\sigma})) + n \sum_{\sigma \in 2^{n}}\int_{[\sigma]} f(s)ds \]

\[=\text{Riemann Sum}[-f(.)\text{log}(f(.)), T , \text{Mesh Size}= 2^{-n}] + n \int_{0}^{1} f(s)ds,\]
where $T=\{x_\sigma : \sigma \in 2^n\}$.
Since the integral in the second term is equal to 1,
\[H(\rho_n)-n = \text{Riemann Sum}[-f(.)\text{log}(f(.)), T, \text{Mesh Size}= 2^{-n}]\]

By assumption, we have that $-\int_{0}^{1}f(s)\text{log}(f(s))ds = c$, for some constant $c$. By standard analysis (See for example, Chapter 3, Section 2 in \cite{PughCharlesChapman2015RMA}), we have that \[\lim_n \text{Riemann Sum}[-f(.)\text{log}(f(.)), T, \text{Mesh Size}= 2^{-n}] = -\int_{0}^{1}f(s)\text{log}(f(s))ds = c.\]
So, for all sufficiently large $n$, we have that $H(\rho_n)-n > c-1 $ as required.
\subsection{The initial segment condition is not necessary for strong quantum randomness}
\label{counter}
We now give an example to show that the converse of Theorem \ref{thm:9} fails. The construction will be along the same lines as in the preceding subsection and we will adopt the notation used there.
Let $f$ be any continuous function on $(0,1)$ satisfying:
\[\int_{0}^{1}f(s)ds =1 \text{ and } \int_{0}^{1}f(s)\text{log}(f(s))ds = \infty.\]
For example, let \[f(x)= \dfrac{1}{x(1-\text{ln} x)^{2}}\] on $(0,1)$.
Define a diagonal state $\rho = (\rho_{n})_{n}$ as follows. Fix $n$. For all $\sigma \in 2^n$, let
\[\alpha_{\sigma} = \int_{[\sigma]} f(s)ds,\]
and let \[\rho_{n}= \sum_{\sigma \in 2^{n}} \alpha_{\sigma} |\sigma\big>\big<\sigma|.\]
$\rho$ is s state since $\alpha_{\sigma}= \alpha_{\sigma 1} + \alpha_{\sigma 0}$ and $\int_{0}^{1}f(s)ds =1$. 

\begin{lem}
The state $\rho$ is strong quantum random.
\end{lem}
\begin{proof}
Let $(G^t)_{t}$ be a quantum null condition. Given an arbitrary $\delta>0$ we show that there is an $k$ such that $\rho(G^k) \leq \delta$. This shows that  $\rho$ is strong quantum random.

Fix an arbitrary $\delta>0$. The function $f$ is in $L_1([0,1], \mu)$, the space of integrable functions over $[0,1]$ equipped with the Lebesgue measure, $\mu$. So by Corollary.3.6 in \cite{folland2013real}, there is a $m$ such that
\[\bigg|\int_{J} f(s)ds\bigg| < \delta,\] 
for any $J\subseteq [0,1]$, with $\mu(J) \leq 2^{-m}$.

Since $\lim_t \tau(G^t)=0$, fix a $k$ such that $\tau(G^k)<2^{-m}$. Suppose that $G^k$ is on $\mathbb{C}^{2^n}$.
We prove that $\rho(G^k)=\tr(\rho_{n}G^{k}_{n})<\delta$. Since $\tau(G^k)<2^{-m}$, $G^{k}$ is a Hermitian projection with rank at most $2^{n-m}$. Let $L$ be the set of $\sigma$'s corresponding to the $2^{n-m}$ largest eigenvalues of $\rho_{n}$. I.e., L = $\{\sigma_{1},\cdots,\sigma_{2^{n-m}}\}$ where the strings are picked such that $\{\alpha_{\sigma_{1}},\cdots,\alpha_{\sigma_{2^{n-m}}}\}$ are the first $2^{n-m}$ many largest eigenvalues of $\rho_n$. By Theorem \ref{thm:10},
\[\tr(\rho_{n}G^{k}_{n})\leq  \sum_{\sigma \in L} \alpha_{\sigma} = \sum_{\sigma \in L} \int_{[\sigma]} f(s)ds =\int_{E} f(s)ds,\] where $E := \bigcup_{L}[\sigma]$ is a disjoint union of open intervals. Since $|L|= 2^{n-m}$ and $\mu([\sigma])= 2^{-n}$ for each $\sigma \in L$, we have that $\mu(E)= 2^{-n}2^{n-m}= 2^{-m}$. So, by the choice of $m$,
\[\bigg|\int_{E} f(s)ds\bigg|<\delta,\] thus showing that $\tr(\rho_n G^k_n)\leq\delta$.
\end{proof} 
We now show that $\rho$ does not satisfy the initial segment condition on von Neumann entropy.
\begin{lem}
For all natural numbers $c$, for almost every $n$, $H(\rho_n) <n-c$.
\end{lem}
\begin{proof}
For any $n$, by definition of the $\alpha$s, we have, \[H(\rho_{n}) = -\sum_{\sigma \in 2^{n}}\int_{[\sigma]} f(s)ds \text{log}(\int_{[\sigma]} f(s)ds)\]
By the mean-value theorem and continuity of $f$, for all $\sigma$ there is a $x_{\sigma}\in [\sigma]$ such that
\[\int_{[\sigma]} f(s)ds = 2^{-n}f(x_{\sigma})\]
So,
\[H(\rho_{n}) = -\sum_{\sigma \in 2^{n}}2^{-n}f(x_{\sigma}) \text{log}(2^{-n}f(x_{\sigma}))\]
\[=-\sum_{\sigma \in 2^{n}}2^{-n}f(x_{\sigma}) (-n+\text{log}(f(x_{\sigma})))\]

\[=-\sum_{\sigma \in 2^{n}}2^{-n}f(x_{\sigma}) \text{log}(f(x_{\sigma})) + n \sum_{\sigma \in 2^{n}}2^{-n}f(x_{\sigma})\]
By definition,
\[=-\sum_{\sigma \in 2^{n}}2^{-n}f(x_{\sigma}) \text{log}(f(x_{\sigma})) + n \sum_{\sigma \in 2^{n}}\int_{[\sigma]} f(s)ds \]

\[=\text{Riemann Sum}[-f(.)\text{log}(f(.),T_n, \text{Mesh Size}= 2^{-n}] + n \int_{0}^{1} f(s)ds. \]
where $T_n=\{x_\sigma : \sigma \in 2^n\}$.
As the integral in the second term is equal to 1,
\[H(\rho_n)-n = \text{Riemann Sum}[-f(.)\text{log}(f(.),T_n, \text{Mesh Size}= 2^{-n}].\]

Recall that \[-\int_{0}^{1}f(s)\text{log}(f(s))ds = -\infty.\]

So, by standard analysis (See for example, Chapter 3, Section 2 in \cite{PughCharlesChapman2015RMA}),

\[\lim_{n\to\infty}H(\rho_n)-n =\lim_{n\to\infty} \text{Riemann Sum}[-f(.)\text{log}(f(.),T_n , \text{Mesh Size}= 2^{-n}]\]
\[= -\int_{0}^{1}f(s)\text{log}(f(s)ds = -\infty.\]

So, for all natural numbers $c$ there is an $N$ such that $n>N$ implies that $H(\rho_n)-n < -c $.

\end{proof}
\section{Uniform integrability and computable quantum random states}
\label{ui}
Uniform integrability is an important real analytic notion used extensively in probability theory \cite{rudin, ShiryaevAlbert2013P}. It is a property of a family of integrable functions defined on a measure space. We specialize this to the setting relevant to us:
\begin{defn}
    
A family, $\mathcal{F}$ of Lebesgue integrable functions on the measure space $([0,1),\mu)$, (the interval $[0,1)$ equipped with the Lebesgue measure, $\mu$) is uniformly integrable if each $f\in \f$ is in $L^1([0,1),\mu)$ and if for all $\delta>0$ there is a $\epsilon>0$ such that \[\mu(S)\leq \epsilon \implies \sup_{f \in \f} \bigg|\int_S f(x)d\mu(x) \bigg| < \delta.\]
(See page 133 in \cite{rudin} and pages 188 to 190 in \cite{ShiryaevAlbert2013P}).
\end{defn}

This section shows that quantum Schnorr randomness of a computable state $\rho=(\rho_n)_n$ is equivalent to uniform integrability of the family of the distributions of the $\rho_n$'s.

The ideas are as follows: Given a state, $\rho=(\rho_n)_n$, we define a family $\f^\rho = (f^\rho_n)_n$ of non-increasing step functions on $[0,1)$ such that $f^\rho_n$ encodes the distribution of $\rho_n$. These functions are constructed such that for any projection $G$ of rank $2^{n-r}$ we have \[\tr(\rho_n G) \leq \int_0^{2^{-r}} f^\rho_n(x) d\mu(x).\] Note that the upper bound on $\tr(\rho_n G)$ \emph{depends only on the rank of} $G$ (This uses Theorem \ref{thm:10}, a corollary of the singular value decomposition). Since quantum Schnorr randomness of $\rho$ depends on the behavior of $\tr(\rho_n G)$ for various computable projections $G$ of small rank, this result can be used to show that the quantum Schnorr randomness of $\rho$ is equivalent to the uniform integrability of $\f^\rho$ for computable $\rho$.

We now describe the construction of $\f^\rho$.

Fix a computable state $\rho =(\rho_n)_n$. For each $n$, let
 \[\rho_{n}= \sum_{i \leq 2^{n}} \alpha^n_{i} |\xi^n_i\big>\big<\xi^n_i|,\] 
 where $(\alpha^n_i, |\xi^n_i\big>)_i$ are the eigenpairs of $\rho_n$. We assume an indexing so that $\alpha^n_1 \geq \cdots \geq \alpha^n_{2^n}$. 

For each $n$, express $[0,1)$ as disjoint a union of $2^n$ many half open intervals $(J^n_i)_{i=1}^{2^n}$, each of length $2^{-n}$. Formally, we have that $J^n_i = [(i-1)2^{-n}, i2^{-n})$ for all $n$ and all $1\leq i\leq 2^n$. 
%I.e., \[J^n_1 = [0, 2^{-n}), J^n_2 = [2^{-n}, 2 * 2^{-n}), J^n_3 = [2*2^{-n}, 3*2^{-n}), \cdots, J^n_{2^n} = [(2^n -1)2^{-n}, 1). \] Not needed

We define a family of functions, $\f^\rho= (f^\rho_n)_n$ as follows:
For each $n$, let $f^\rho_n$ be the function on $[0,1)$ defined by: 
\[f^\rho_n (x):= 2^n \alpha^n_i  ,\text{  if }  x \in J^n_i. \]
For each $n$, the function $f^\rho_n$ is well defined since $[0,1) = \bigcup_{i \leq 2^n} J^n_i$ and this is a disjoint union. Note also that $f^\rho_n$ is non-increasing since $\alpha^n_1 \geq \cdots \geq \alpha^n_{2^n}$. This will be important later.
Note also that
\[\int_0^1 f^\rho_n (x) d\mu(x) = \sum_\sigma \int_{[\sigma]} f^\rho_n (x) d\mu(x) = \sum_\sigma 2^n \alpha^n_\sigma \mu([\sigma])= \sum_\sigma \alpha^n_\sigma = 1,\]
showing that each member of $f^\rho$ is in $L_1([0,1],\mu)$. We now show the main result of this section:
\begin{thm}
\label{uli}

 A computable state $\rho$ is quantum Schnorr random iff $\f^\rho$ is uniformly integrable.
\end{thm}
This will follow from Lemmas \ref{req} and \ref{unifstrong}. Lemma \ref{req} which shows that quantum Schnorr randomness of $\rho$ implies uniform integrability of $\f^\rho$ holds for computable $\rho$. Lemma.\ref{unifstrong} shows the stronger result that uniform integrability of $\f^\rho$ implies strong quantum randomness for any $\rho$. This suggests that the equivalence in Theorem.\ref{uli} only holds for computable $\rho$ and that uniform integrability of $\f^\rho$ may be a stronger condition than quantum Schnorr randomness for non computable states.

\begin{lem}
\label{req}
 If a computable state $\rho$ is quantum Schnorr random then $\f^\rho$ is uniformly integrable. 
\end{lem}
\begin{proof} 

    Let $\rho=(\rho_n)_n$ be a computable quantum Schnorr random state. For a contradiction, let $\f^\rho$ not be uniformly integrable. So, there is a $\delta>0$ such that for all $m$, there exists a $n(m)$ and a set $W_m$ such that $\mu(W_m)\leq 2^{-m}$ and \[\int_{W_m} f^\rho_{n(m)}(x) d\mu(x) >\delta.\]
    Fix an $m$ and let the corresponding $n(m)$ be denoted by $n$.
    Since $f^\rho_n$ is non-increasing and $\mu(W_m)\leq 2^{-m}$, is it easy to see that

   \[\int_{W_m} f^\rho_n(x)d\mu(x)\leq \int_0^{2^{-m}} f^\rho_n(x)d\mu(x). \]
Noting that $[0,2^{-m}) = \bigcup_{i\leq 2^{n-m}} J^n_i$ and using the above property of $W_m$,

  \[\delta< \int_0^{2^{-m}} f^\rho_n(x)d\mu(x) = \sum_{i\leq 2^{n-m}} \int_{J^n_i} f^\rho_n(x)d\mu(x) = \sum_{i\leq 2^{n-m}} \alpha^n_i.\]
  
We define a q-S test, $(G^m)_m$:  Given an arbitrary $m$, compute a $j$ such that, $\sum_{s\leq 2^{j-m}} \alpha^j_s$, the sum of the first $2^{j-m}$ many largest eigenvalues of $\rho_j$ exceeds $\delta$. Let \[G^m := \sum_{i\leq 2^{j-m}} |\xi^j_{_i}\rangle\langle \xi^j_{_i}|.\]

 We verify that $(G^m)_m$ is a q-S test and that $\rho$ fails it at order $\delta$.

By the above argument, for each $m$, there is a $n$ such that $\delta< \sum_{i\leq 2^{n-m}} \alpha^n_i$. So, the $j$ needed to define $G^m$ exists. Further, since $\rho$ is computable, the eigendecomposition of each $\rho_n$ is uniformly computable in $n$ and thus this $j$ is computable uniformly in $m$. The construction of $G^m$ is thus uniform in $m$.
Note that $\tr(G^m)= 2^{j-m}$. Since $G^m$ is on $\mathbb{C}^{2^j}$ we have $\tau(G^m)=2^{-m}$. This shows, by Remark \ref{schnorr}, that $(G^m)_m$ is a q-S test. 

Fix an $m$ and let $j$ be as above. Then, \[ \rho(G^m)\geq \tr(\rho_{j} G^m) = \sum_{i\leq 2^{j-m}}\langle \xi^j_{i}|\rho_j  |\xi^j_{i}\rangle = \sum_{i\leq 2^{j-m}}\alpha^j_{i}> \delta.\]
The last inequality follows by the choice of $j$. As $m$ was arbitrary, we see that $\rho(G^m)>\delta$ for all $m$ and hence that $\rho$ fails $(G^m)_m$ at order $\delta$. This is a contradiction.
\end{proof}

\begin{lem}
\label{unifstrong}
    Let $\rho$ be any state. If $\f^\rho$ is uniformly integrable, then $\rho$ is strongly quantum random.
\end{lem}
\begin{proof}

Let $\rho=(\rho_n)_n$ be a state such that $\f^\rho$ is uniformly integrable. Suppose for a contradiction that $\rho$ fails to satisfy a quantum null condition $(G^j)_j$. So there is a $\delta>0$ such that $\rho(G^j)>\delta$ for all $j$.

By uniform integrability, fix a $m$ such that if $\mu(W)\leq 2^{-m}$, then $\int_W f^\rho_n (x)d\mu(x) <\delta$ for all $n$.

Let $G^q$ be such that $\tau(G^q)<2^{-m}$. Fix the $k$ such that $G^q$ is on $\mathbb{C}^{2^k}$. So rank$(G^q)=\tr(G^q) < 2^{k-m}$ (Recall Definition \ref{defn:sigclass}). By Theorem \ref{thm:10} and by our assumption,  \[\sum_{i\leq 2^{k-m}} \alpha^k_{i} \geq \tr(\rho_k G^q) = \rho(G^q) > \delta.\]

Also, since $[0,2^{-m})=\bigcup_{i\leq 2^{k-m}} J^k_i,$

\[\int_0^{2^{-m}} f^\rho_k (x) d\mu(x) = \sum_i \int_{J^k_i} f^\rho_k (x) d\mu(x) = \sum_i 2^k \alpha^k_{i} \mu(J^k_i)= \sum_{i\leq 2^{k-m}} \alpha^k_{i} > \delta.\]
This contradicts the choice of $m$ since $\mu((0,2^{-m}])=2^{-m}$.
\end{proof}

\section{Discussion and open questions}

Our main results are that the quantum Schnorr randomness of a computable $\rho$ strictly implies $H(\rho)=1$ (Section \ref{liminf}) and that for any $\rho$, $\exists c \exists^\infty n, H(\rho_n)>n-c$ strictly implies that $\rho$ is strong quantum random (Section \ref{ls}).  These results lead us to ask:

\begin{oq}
   
Can we characterize quantum Schnorr randomness for computable states using initial segment von Neumann entropy? 
\end{oq}
Uniform integrability of the distributions of a state's initial segments is used to characterize quantum Schnorr randomness for computable states in Section \ref{ui}.
Quantum s-tests are used to give a finer-grained relation between the entropy rate and randomness of computable states: $H(\rho)\geq \inf\{s: \rho$ is covered by a q s-test$\}$ for computable $\rho$ (Section \ref{rqr}). The inequality is an equality for the states in Definition \ref{power}. This naturally leads us to ask:
\begin{oq}
\label{open}
    Is it true that $H(\rho)=\inf\{s: \rho$ is covered by a q s-test$\}$ for any computable $\rho$? 
\end{oq}

Lemma \ref{power} which answers this for states in Definition \ref{defnd} uses the quantum typical subspace theorem which applies to a fixed density matrix. As there is no analogue of this theorem for an infinite sequence of density matrices, the method of Lemma \ref{power} does not seem to extend to general states.

We note a stark difference between classical and quantum algorithmic randomness. If a $Y \in 2^\omega$ is computable, then it cannot be Schnorr random since each $Y\upharpoonright n $ is computable and hence has low descriptive complexity. By contrast, there are computable quantum Schnorr random states. This disparity seems to reflect the replacement of bitstrings by density matrices when transitioning from the classical to the quantum setting. While the randomness of an infinite bitrsing is \emph{entirely} characterized by the descriptive complexities of its initial segments, the randomness of a state has two aspects:

\begin{enumerate}
    \item The descriptive complexity of the eigenvectors of its initial segments.
    \item The uniformity (entropy) of the distributions of its initial segments. 
\end{enumerate}

A state whose initial segments have eigenvectors with low descriptive complexity and highly uniform distributions is random since its initial segments cannot have large projections onto small dimensional subspaces. The randomness of such a state stems from (2). On the other hand, a state whose initial segments have eigenvectors with high descriptive complexity and non-uniform distributions will be random since although the distribution peaks at few eigenvectors, these eigenvectors cannot be computably described and hence cannot be used to construct a test. The randomness of such a state stems from (1). Since the eigenvectors of the initial segments of a computable state have low descriptive complexity, the randomness of computable states necessarily stems from (2). The results in Sections.\ref{liminf},\ref{rqr} and \ref{ui} apply only to computable states hence capture aspect (2).

Nies and Stephan have defined Martin-L{\"o}f absolute continuity (ML.a.c); a notion of randomness for measures on Cantor space\cite{NIES20221}. A measure $\nu$ is ML a.c. if $\nu(\MLR)=1$ where $\MLR$ is the set of Martin-L{\"o}f random reals. Any measure $\nu$ on Cantor space induces a state, $\rho^\nu = (\rho^\nu_n)_n$ defined by
$\rho^\nu_n = \sum_{\xi \in 2^n} \nu([\xi])|\xi\rangle \langle\xi|,$
where $\{|\xi\rangle : \xi \in 2^n\}$ is the computational basis of $\C^{2^n}$\cite{unpublished}. The von Neumann entropy of the length $n$ initial segments of $\rho^\nu$ is simply the Shannon entropy of $\nu_n$, which is the restriction of $\nu$ to $2^n$. It is known that $\nu$ is ML.a.c iff $\rho^\nu$ is quantum Martin-L{\"o}f random \cite{unpublished, NIES20221}. Section \ref{liminf} and Remark \ref{mlrtoo} gives that any computable ML.a.c measure $\nu$ has $\lim_n H(\nu_n)/n = 1$. This implication is strict since the example in Lemma \ref{ex} is a measure on Cantor space. Section.\ref{ls} gives that if $\exists c \exists^\infty H(\nu_n)>n-c$ then $\rho^\nu$ is strong quantum random from which it easily follows that $\nu$ is ML a.c with respect to any oracle. Being ML a.c for any oracle simply means that $\nu$ is absolutely continuous with respect to $\mu$, the Lebesgue measure on $[0,1)$. 
We thus have a condition on the Shannon entropies of the $\nu_n$'s which implies absolute continuity of $\nu$ with respect to $\mu$.
This implication is strict since the counterexample to it constructed in Section \ref{counter} is a measure on Cantor space.  
\section{Acknowledgements}
The paper was written during my time at the University of Wisconsin-Madison as a PhD student under Joseph S. Miller and during my time as a post-doctoral fellow at the University of Auckland under Andr\'e Nies. The latter stint was funded by the Marsden grant UOA1931 awarded by the RSNZ to Professor Nies. I am grateful to both JM and AN for numerous discussions.
I am indebted to AN for thoroughly reviewing the paper and for suggesting several significant improvements to it.

\bibliographystyle{eptcs}
\bibliography{references}

\end{document}